\documentclass[fleqn,usenatbib,useAMS]{mnras}

\usepackage{newtxtext,newtxmath}
\usepackage[T1]{fontenc}
\usepackage{ae,aecompl}

\usepackage{graphicx}	% Including figure files
\usepackage{amsmath}	% Advanced maths commands
\usepackage{amssymb}	% Extra maths symbols

\title[The evolution of Giant Molecular Filaments]{The evolution of Giant Molecular Filaments}

\author[Duarte-Cabral et al.]{
Ana Duarte-Cabral$^{1,2}$\thanks{E-mail: adc@astro.cf.ac.uk}, 
and C. L. Dobbs$^{2}$ \\
$^{1}$School of Physics \& Astronomy, Cardiff University, Queen's building, The parade, Cardiff, CF24 3AA, U.K. \\
$^{2}$School of Physics, University of Exeter, Stocker Road, Exeter, EX4 4QL, U.K. \\
}

\date{Accepted 2017 June 15. Received 2017 June 15; in original form 2017 March 1}

\pubyear{2017}

\begin{document}
\label{firstpage}
\pagerange{\pageref{firstpage}--\pageref{lastpage}}
\maketitle

\begin{abstract}

In recent years there has been a growing interest in studying giant molecular filaments (GMFs), which are extremely elongated ($>$ 100\,pc in length) giant molecular clouds (GMCs). They are often seen as inter-arm features in external spiral galaxies, but have been tentatively associated with spiral arms when viewed in the Milky Way.
In this paper, we study the time evolution of GMFs in a high-resolution section of a spiral galaxy simulation, and their link with spiral arm GMCs and star formation, over a period of 11\,Myrs.  The GMFs generally survive the inter-arm passage, although they are subject to a number of processes (e.g. star formation, stellar feedback and differential rotation) which can break the giant filamentary structure into smaller sections. The GMFs are not gravitationally bound clouds as a whole, but are, to some extent, confined by external pressure. Once they reach the spiral arms, the GMFs tend to evolve into more substructured spiral arm GMCs, suggesting that GMFs may be precursors to arm GMCs. Here, they become incorporated into the more complex and almost continuum molecular medium that makes up the gaseous spiral arm. Instead of retaining a clear filamentary shape, their shapes are distorted both by their climb up the spiral potential and their interaction with the gas within the spiral arm. The GMFs do tend to become aligned with the spiral arms just before they enter them (when they reach the minimum of the spiral potential), which could account for the observations of GMFs in the Milky Way.  

\end{abstract}


%%%%%%%%%%%%%%%%%%%%%%%%%%%%%%%%%%%%%%%%%%%%%%%%%%

%%%%%%%%%%%%%%%%% BODY OF PAPER %%%%%%%%%%%%%%%%%%

\section{Introduction}

Observations of nearby grand-design spiral galaxies have long been fascinating in revealing the distribution of dark lanes of dense gas and intense regions of star-formation, across spiral arms and inter-arm regions \citep[e.g.][]{Elmegreen1980,Elmegreen1989,Scoville2001,LaVigne2006,Corder2008,Schinnerer2013,Schinnerer2017}. Though limited in their ability to probe and resolve the detailed structures of the interstellar medium, those studies have provided us with insightful clues concerning the global structure of our very own Milky Way \citep[][]{Benjamin2008}, for which the line-of-sight confusion inherent of an edge on perspective is a constant limitation. 

In recent years, with a growing knowledge of the global structure of our Galaxy and the wealth of high-resolution IR and sub-mm surveys across the Galactic plane, it has become possible to start to unravel the detailed morphology of the molecular gas structures in our Galaxy, and their link to the larger-scale Galactic environment. 
In particular, after the discovery of the striking $\sim 80$\,pc long ``Nessie'' filament \citep{Jackson2010} in the Galactic plane, a number of other observational studies \citep[e.g.][]{Beuther2011,Li2013,Goodman2014,Ragan2014,Zucker2015,Wang2015,Wang2016,Abreu-Vicente2016} have since tried to find other Nessie-like filaments in the Milky Way, other `giant molecular filaments' (GMFs), and associate them with the Galactic large-scale spiral structure.

We now have a sample of about $\sim$30 GMFs throughout the Galaxy (with lengths of the range $\sim 40 - 500$\,pc). The majority of observational studies tend to favour associating these GMFs with spiral arms. However, these results are very susceptible to the particular spiral arm model taken to specify the position and extent of the Galactic spiral arms, whose exact number, location, and shape are still a current challenge to define \citep[e.g.][]{sewilo2004,Benjamin2008,Francis2012,Hou2015}. Furthermore, the distances (and de-projection) of these GMFs onto the Galactic plane are based on the kinematical information across these filaments, which comes with strong uncertainties, in particular around spiral arms, where the velocities can deviate from the circular motions of the stars. 

Another way to understand the nature and origin of such giant filamentary clouds is to use numerical simulations of spiral galaxies, which track the life-cycle of gas over several orbits. Highly elongated arm-related sub-structures have been found in numerical simulations of galaxies \citep[e.g.][]{Kim2002,Chak2003,Shetty06,DobbsBonnell2006,Wada2011,WangHH2015}, often referred to as spurs or feathers. These are typically elongated clouds that are in the process of exiting a spiral arm (i.e. downstream of the spiral wave). Spurs are therefore the remnants of spiral arm clouds, re-entering the shear-dominated inter-arm region.  As they travel in the inter-arms, they are stretched into long filamentary clouds and sometimes even ``bridges'' that connect consecutive spiral arms. However, these simulations did not have sufficient resolution to address the link between spurs and star formation (SF), and molecular cloud formation in spiral arms. 
 
In addition to GMFs in our Galaxy, spurs or feathers have been observed in M51 \citep[e.g.,][]{Corder2008,Koda2009}, as well as other spiral galaxies \citep[e.g.,][]{LaVigne2006}. These features span the inter-arm regions similar to those in the simulations. However, so far there has been little comparison or discussion regarding how these features relate to the GMFs of our Galaxy.
In this paper, we study a high resolution simulation of gas entering a spiral arm \citep[from][]{Dobbs2015}, and the evolution of GMFs. Since the simulation used initial conditions from a galaxy simulation which had already evolved, the GMFs have already formed through the shearing of giant molecular clouds (GMCs) as they leave the spiral arm. Here we focus on their evolution, and associated star formation and gas content, as they cross the inter-arm region, and re-enter a spiral arm.

\section{Method}
\subsection{The numerical model}

In this paper, we study the time evolution of clouds within a Smoothed Particle Hydrodynamics (SPH) simulation similar to that described in \citet{dc2016}, although now with the inclusion of star particles, which mimic the formed star clusters. The specific model we use here is described in \citet{Dobbs2015} as Run 5. This is a section of the galaxy model presented in \citet{Dobbs2013} simulated at higher-resolution, with a particle mass of $\sim\,3.85$\,M$_{\odot}$. As explained in  \citet{Dobbs2015}, we extracted a region of gas by selecting a 1\,kpc by 1\,kpc box along a spiral arm, from the full galaxy simulation of \citet{Dobbs2013}, and then tracing those gas particles back in time by 50\,Myr. In addition, we also included neighbouring particles, so that gas interactions with neighbours are still included for the timescales of the resimulation. This model includes self-gravity, heating and cooling, and simple H$_{2}$ and CO formation \citep{Dobbs08,Pettitt2014}. The minimum temperature of gas in the simulation is 50\,K. This simulation also includes the galactic disc potential as well as a two-armed spiral potential, as in the original simulation from \citet{Dobbs2013}. Thus the gas still rotates around the galaxy, and feels the same external gravitational potential as in the large galaxy-scale simulations. We assume that star formation occurs whenever gas lying above a 500\,cm$^{-3}$ density threshold is both bound and converging. Star particles are formed for each star formation event \citep[as described in][]{Dobbs2014}, and only experience gravity, but not pressure. After the star particles are formed, stellar feedback is included using the same method as in \citet{Dobbs11b}, where feedback is inserted using a stochastic prescription, and the feedback energy is continuously inserted into the gas surrounding the star particle over a period of 5\,Myr (as opposed to being instantaneous), with a feedback efficiency of $\epsilon=0.15$. The amount of energy inserted is 10$^{51}$\,ergs per massive star formed, and we assume that one massive star forms per 160\,M$_{\odot}$ of stars \citep[see equation 1 and accompanying text in][]{Dobbs2015}. Although the feedback uses a supernova snowplough solution to determine the (kinetic and thermal) energy to be inserted into the gas, the continuous injection of energy can also be supposed to represent winds and ionisation that precede a supernova. A more detailed description of this simulation can be found in \citet{Dobbs2015}. In this paper, we analyse this model over a total period of 11\,Myrs, from 9\,Myrs to 20\,Myrs, at intervals of $\sim$1\,Myr.

\subsection{Following clouds over time}

\begin{figure*}
\centering
\hspace{2cm}\includegraphics[width=0.85\textwidth]{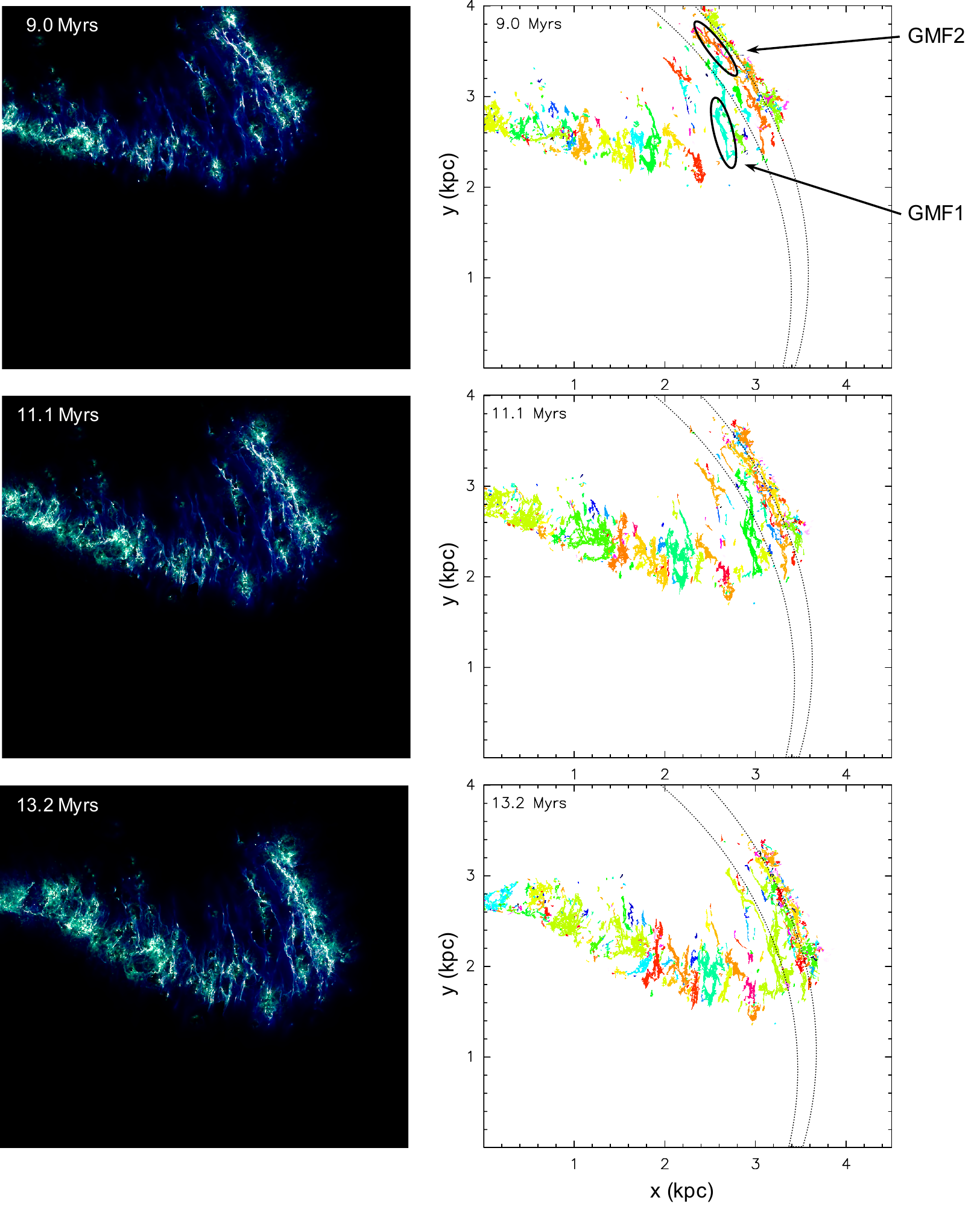}
\vspace{-0.4cm}
\caption{\small{Time evolution of the simulation used in this work (converted onto 5\,pc grid-size datacubes), where each row is a time step separated by 2\,Myrs, as labelled on the top-left corner of each panel. {\it Left:} Top-down view of the simulation as 3-colour (RGB) images of the column densities of CO (red), H$_{2}$ (green) and atomic H (blue). {\it Right:} Mask of the corresponding GMCs extracted from the H$_{2}$ density datacube with the {\sc scimes} code, where each cloud is represented by one colour, which relates to the cloud's centroid $z$ coordinate (blue being lower $z$ values and red being higher $z$). The dotted lines outline the position of the minimum of the spiral potential well (at $\psi_{sp}=-380$~km$^2$~s$^{-2}$). The two labelled ellipses on the top-right panel show the position of the two GMFs we track at higher resolution.}}
\label{fig:top-down-molecular}
\vspace{-0.3cm}
\end{figure*}

\begin{figure*}
\centering
\hspace{2cm}\includegraphics[width=0.85\textwidth]{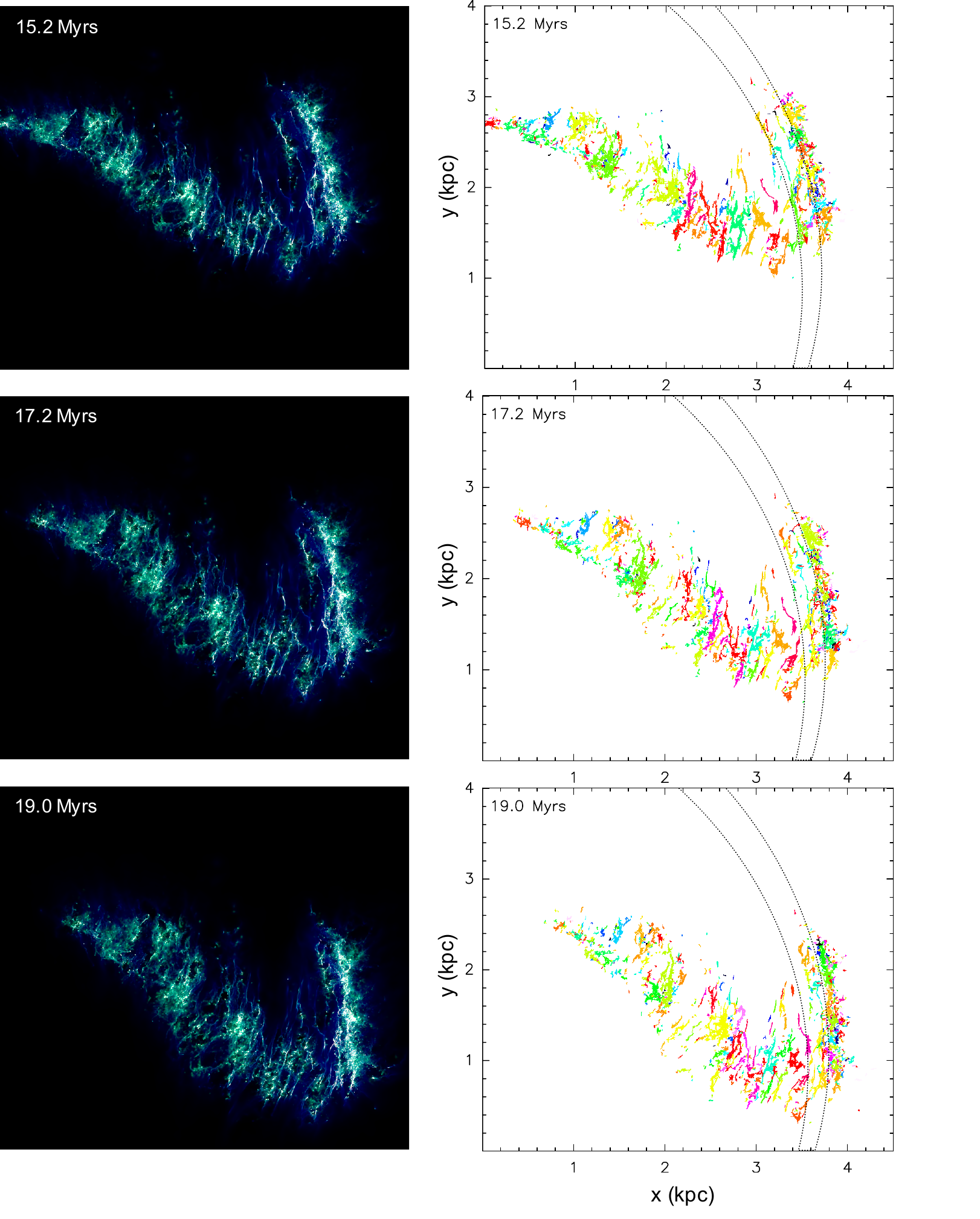}
\vspace{-0.4cm}
\caption{\small{Fig.~\ref{fig:top-down-molecular}, continued.}}
\vspace{-0.3cm}
\label{fig:top-down-molecular2}
\end{figure*}

To study the population of molecular clouds over time, and in order to compare these time-evolution results with those obtained by \cite{dc2016} from a single snapshot, we have used the same method as in \cite{dc2016} to extract clouds at each simulation timestep, using {\sc scimes} \citep[Spectral Clustering for Interstellar Molecular Emission Segmentation, see][for full details]{Colombo15}, which is a code designed to identify GMCs in observations based on cluster analysis. As this code works on grid-based datacubes, we have built three-dimensional datacubes of the entire simulation, with a regularly spaced grid of 5\,pc in size, with the volume densities of H, H$_{2}$ and CO, as extracted from the SPH data with {\sc splash} \citep{Price2007}. The resulting datacubes can be seen in Figures\,\ref{fig:top-down-molecular} and \ref{fig:top-down-molecular2}.
 
The extraction of GMCs was made on the H$_{2}$ density cubes, as we are interested in following the evolution of the large molecular gas complexes, even if a fraction of these may not have observable levels of CO. As noted in \cite{dc2016}, although a 5\,pc grid resolution is enough to determine the overall distribution of clouds, it is relatively coarse in order to study the properties of giant filaments, whose widths are generally unresolved. We have thus selected two of the largest GMFs in the simulation to track at higher resolution, for which we built datacubes of 1\,pc grid resolution. We re-extracted the clouds on the higher-resolution datacubes using the same extraction algorithm ({\sc scimes}), and cross-matched the GMFs between timesteps (as well as their subsequent cloud fragments) by eye.

%==========================

\section{Time evolution of giant molecular filaments}
\label{sec:time_ev}

\subsection{Shaping filaments on a global picture}
\label{sec:global_ev}

\begin{figure*}
\includegraphics[width=\textwidth]{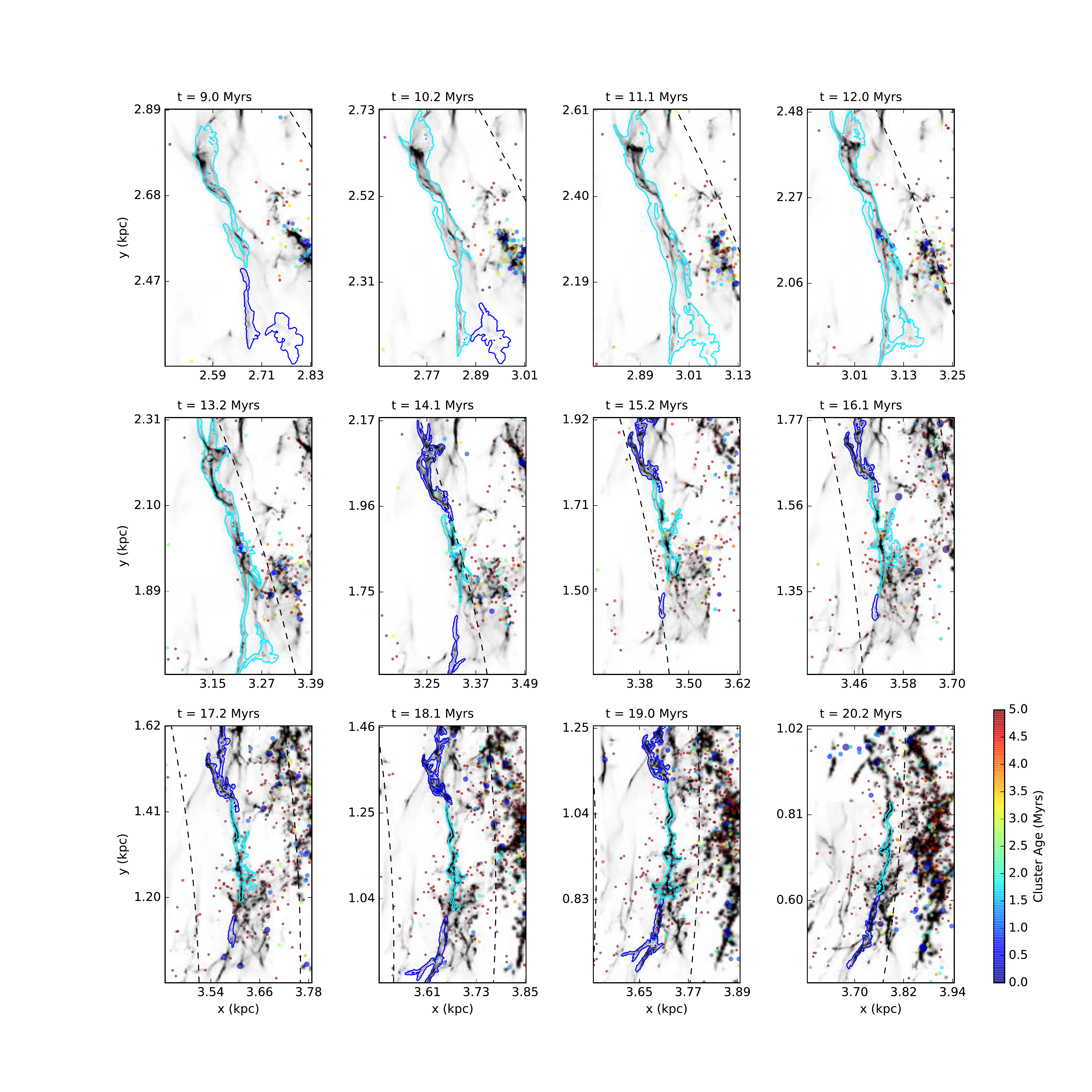}
\vspace{-0.5cm}
\caption{\small{Time evolution of GMF1, where the background grey scale shows the top-down view of the total column density, the cyan contours highlight the main filament, while the dark blue contours follow some of the smaller clouds that were or will be once part of the main GMF. Circles show the population of star particles (representing stellar clusters) size- and colour-coded by their age (large/blue being young, and small/red being 5\,Myrs or older). The dashed lines delineate the bottom of the potential well, as in Fig.\,\ref{fig:top-down-molecular}. The only significant star formation event in GMF1 occurs at $t=12$\,Myrs, which results in a visible shell on the following timesteps. At the end of the run, GMF1 is still close to the bottom of the potential well, and has yet to enter the gaseous part of the spiral arm (where there is a large concentration of both stellar clusters and gas).}}
\label{fig:evolution_cloudA}
\vspace{-0.2cm}
\end{figure*}

\begin{figure*}
\includegraphics[width=\textwidth]{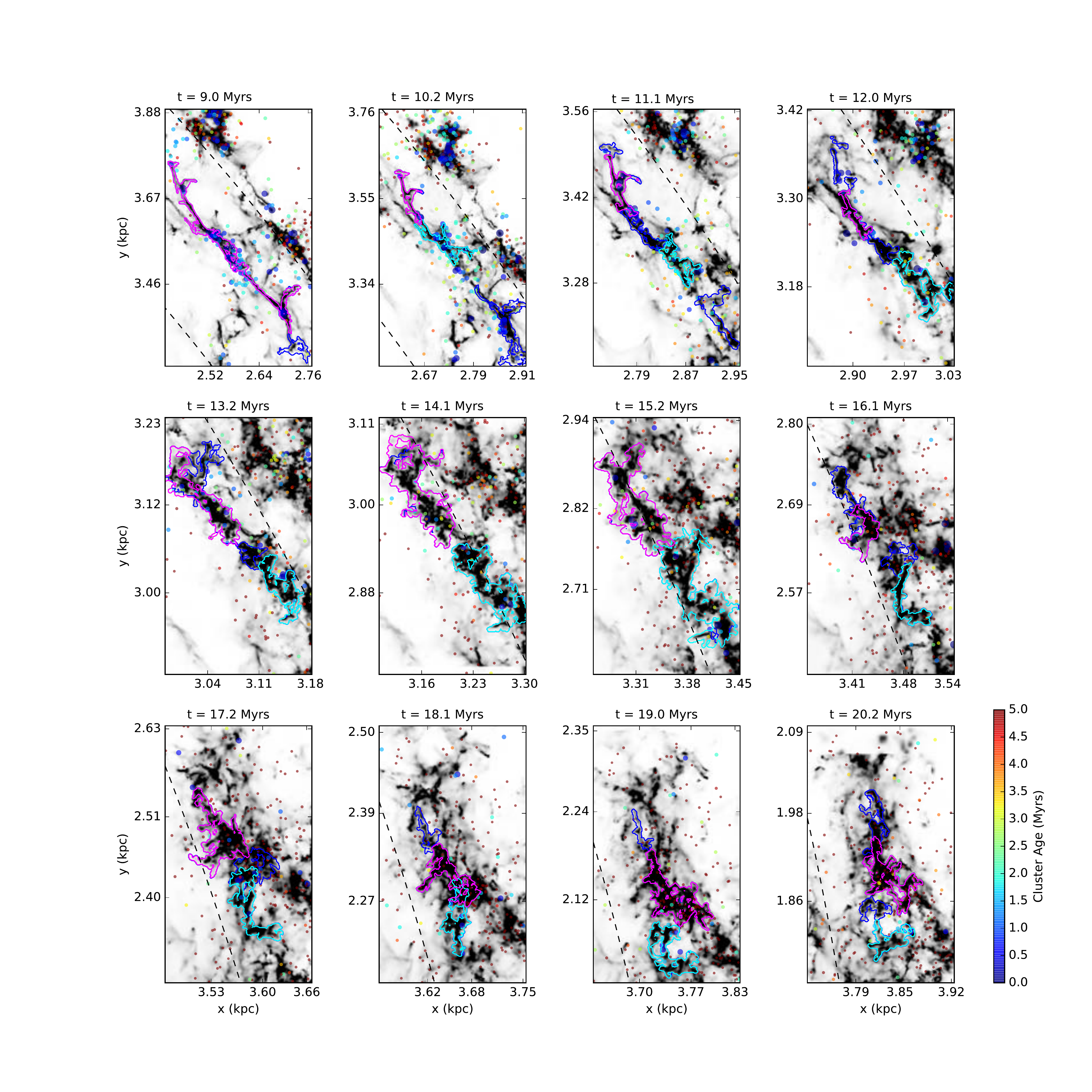}
\vspace{-0.5cm}
\caption{\small{Time evolution of GMF2, where the background grey scale shows the top-down view of the total column density, the magenta and cyan contours highlight the two main filaments, while the dark blue contours follow some of the smaller clouds that were or will be once part of the main GMF. As in Fig.\,\ref{fig:evolution_cloudA}, circles show the population of star particles (representing stellar clusters), and the dashed lines delineate the bottom of the potential well. GMF2 starts close to the bottom of the potential well, and has significant star formation at $t=9-12$\,Myrs, which results in visible shells on the following timesteps, that manage to break up the cloud. From $t=15.2$\,Myrs, GMF2 becomes part of the gaseous spiral arm, merging with other clouds. GMF2 develops a complex morphology, no longer retaining its large-scale filamentary appearance.}}
\label{fig:evolution_cloudG}
\end{figure*}

In Figures\,\ref{fig:top-down-molecular} and \ref{fig:top-down-molecular2} we show the time evolution of the simulation in intervals of 2\,Myr, as well as the respective extraction of GMCs on the coarser 5\,pc grid resolution. The statistical properties of clouds at each timestep (including mass, velocity dispersion, size, aspect ratio, average surface density and virial parameter, see App.\,\ref{app:time_evolution}), remain virtually unchanged over time, with distributions of similar shape and median values (see Table\,\ref{tab:stat_properties}, and Fig.\,\ref{fig:histograms}), suggesting that the results from \cite{dc2016}, for a single timestep, are not time-sensitive.  Still, by examining how clouds evolve over time, we can see more clearly how the large number of very elongated structures are formed and shaped (see Figs.\,\ref{fig:top-down-molecular} and \ref{fig:top-down-molecular2}). These long filaments are particularly striking in the inter-arm regions after crossing the peak of the spiral potential, and approach the spiral potential minimum\footnote{Note that the spiral potential minimum is slightly shifted with respect to the position of the gaseous spiral arm (as defined by the higher density of gas), due to the specific response of the gas to the spiral potential \citep[see also][]{Pettitt2014,Hou2015}}. Some of these GMFs can in fact span $>500$\,pc in length in H$_{2}$, but as is clear from the left panels of Figs.\,\ref{fig:top-down-molecular} and \ref{fig:top-down-molecular2}, the CO (seen as white-pink areas) is only present on the denser ridges of these clouds, splitting these GMFs into smaller filaments of the order of $\sim$100\,pc in length. 

In accordance with the results from \cite{dc2016}, we find that giant molecular filaments\footnote{GMFs were identified by selecting the sample of clouds from our 5\,pc-grid cubes with an aspect ratio larger than 6, a major axis of at least 30\,pc, and a minor axis of less than 15\,pc. The aspect ratio is estimated using the major and minor axes from the ellipsoidal fit of the 3D density structure. However, long filaments can often be curved and be part of networks of filaments, in which case the ellipsoidal fit will provide an artificially low aspect ratio, resulting in our automatic selection potentially missing some GMFs.} are exclusively found in inter-arm regions. The total molecular mass in GMFs varies from a few percent up to 10$\%$ of the total H$_{2}$ mass in GMCs, although this is likely a lower limit due to our selection caveats. In terms of global properties, GMFs do not form an isolated type of clouds, they are simply part of a continuous tail of distributions in sizes and aspect ratio. The ranges of masses, velocity dispersions and virial parameters of the GMFs are similar to the bulk of the clouds.

As can be seen on the right panels of Figs.\,\ref{fig:top-down-molecular} and \ref{fig:top-down-molecular2}, in the gaseous spiral arm we no longer find well defined GMFs. Instead, we find large complexes of clouds, similarly to \cite{dc2016}, forming a near continuum of material in H$_{2}$. This suggests that GMFs incorporate the arms' GMC complexes, and may have difficulty surviving the passage of the spiral arm as coherent structures.

\subsection{Tracking individual giant filaments}
\label{sec:indiv_ev}

\begin{figure*}
\includegraphics[width=0.77\textwidth]{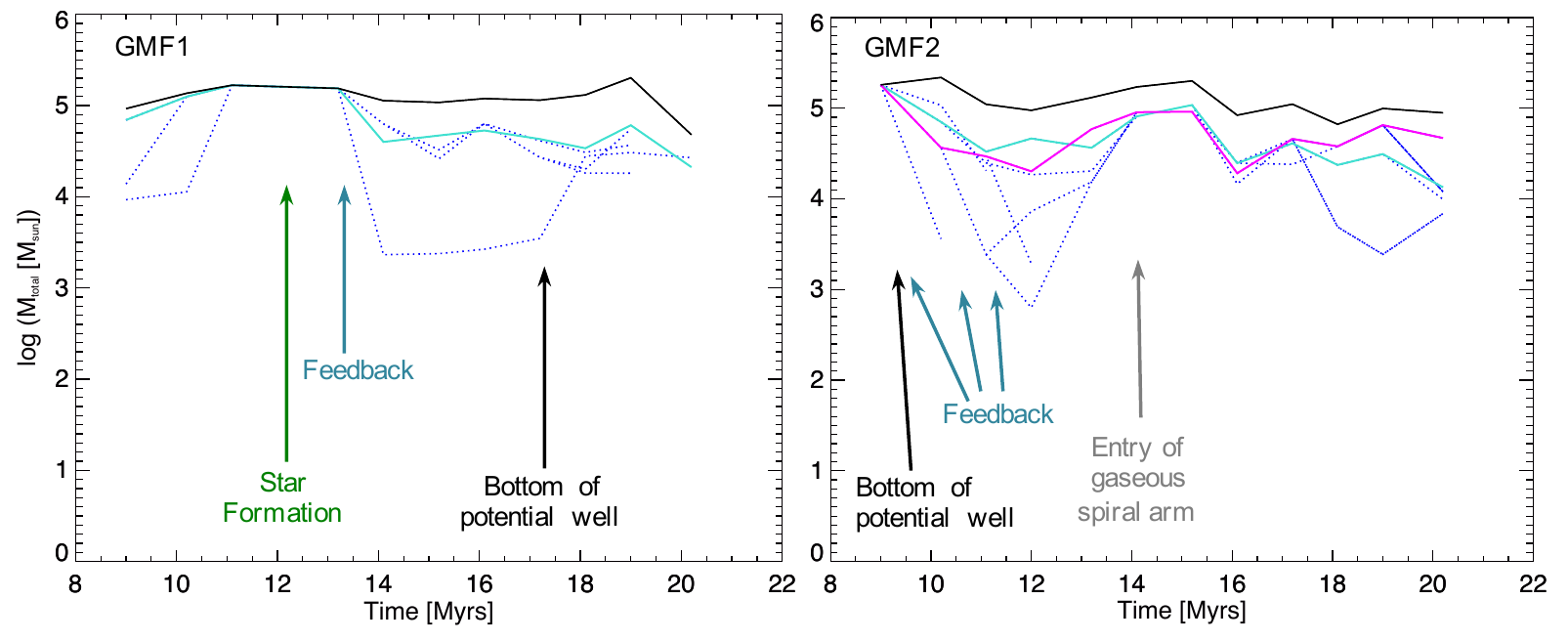}
\vspace{-0.3cm}
\caption{\small{Time evolution of the total mass of GMF1 (left) and GMF2 (right). As the giant filaments evolve, they do not survive as a single entity but are are made up of smaller clouds as the main filament breaks up. The black solid line shows the evolution for the entire cloud (i.e. the total mass within the several fragments), which remains relatively constant over time. The coloured lines show the evolution of different cloud fragments, following the same colour scheme as Figs.\,\ref{fig:evolution_cloudA} and \ref{fig:evolution_cloudG}: solid turquoise line for the main GMF1, solid turquoise and solid magenta for the two main filaments of GMF2; and blue dashed lines for all the remaining smaller individual clouds that were once part of the GMFs.}}
\label{fig:total_mass_evolution}
\end{figure*}

\begin{figure*}
\vspace{-0.2cm}
\includegraphics[width=0.78\textwidth]{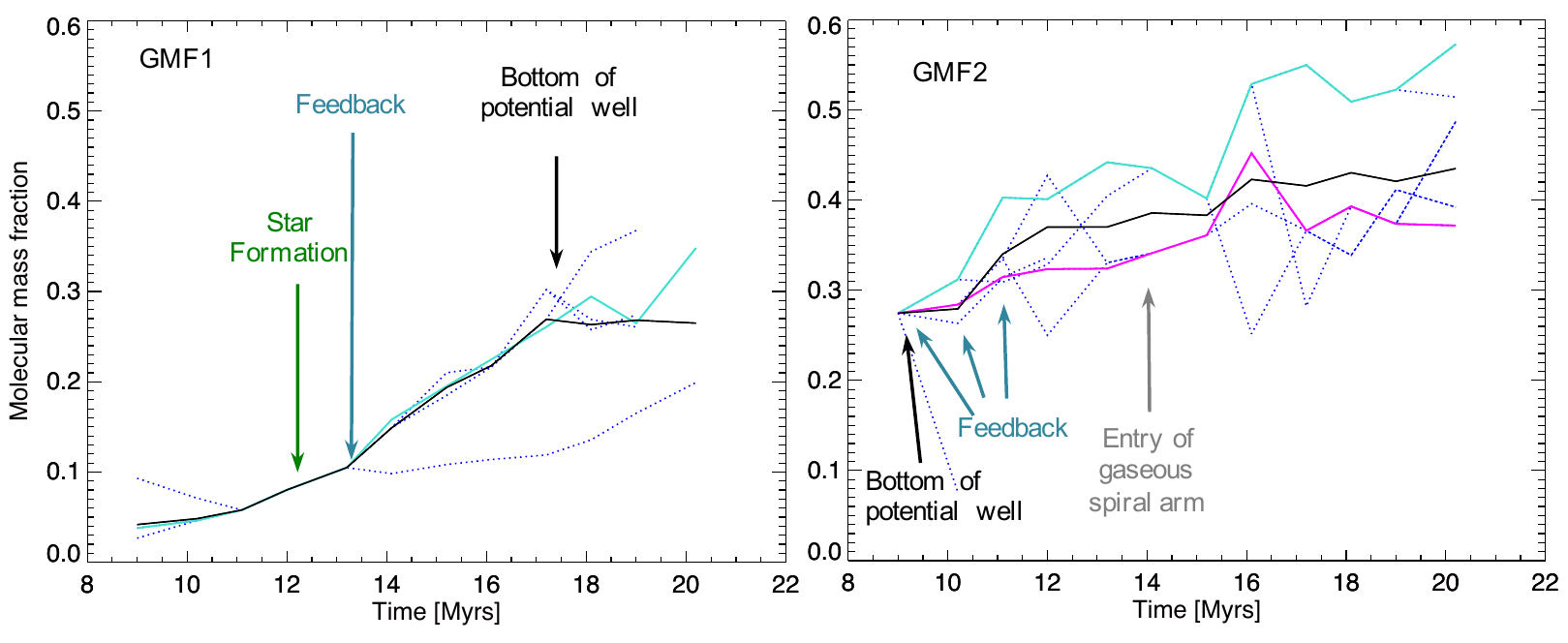}
\vspace{-0.3cm}
\caption{\small{Time evolution of the molecular gas mass fraction of GMF1 (left) and GMF2 (right). Solid and dashed lines are as in Fig.\ref{fig:total_mass_evolution}. Although the total mass of the clouds remains relatively constant over time, the molecular gas fraction increases with time, especially whilst the clouds are being formed in the inter-arm (which is the case of GMF1). The molecular mass fraction of GMF2 also increases, but GMF2 does not reach more than 60\% molecular due to the input of stellar feedback in the denser regions as soon as SF takes place.}}
\label{fig:molecular_fraction_evolution}
\end{figure*}

To study the evolution of giant filaments in more detail, in particular their morphological and chemical evolution, we have selected two representative GMFs at $t=9$\,Myrs, and tracked them over time (for 11\,Myrs), at higher resolution (1\,pc). For this purpose, we have selected one of the longest molecular filaments in the sample, spanning $\sim$600\,pc, located in the inter-arm region (GMF1, see Fig.\,\ref{fig:evolution_cloudA}); and one filament that spans $\sim$500\,pc which is at the bottom of the potential well at $t=9$\,Myrs, and then goes on to enter the gaseous spiral arm (GMF2, see Fig.\,\ref{fig:evolution_cloudG}). In Fig.\,\ref{fig:top-down-molecular}, we show the position and extent of these two GMFs at $t=9$\,Myrs.

In terms of morphology, both GMFs are most defined when they reach the bottom of the spiral potential well (see GMF1 at $t=18$\,Myrs and GMF2 at $t=9$\,Myrs). In effect, as the gas is accelerated towards the bottom of the spiral potential, the density contrast of the ridge of the filaments increases, as a consequence of the gas feeling a ``frontal wind'', or ram pressure. This effectively works as a compression force acting on the smaller axis of these filaments, whilst the galactic shear from the differential rotation, continues to stretch the clouds in their longer axis (see Fig.\,\ref{fig:evolution_cloudA}). This shear is such that the star formation activity, although it does occur, is not particularly strong in the moments of approach of the bottom of the spiral potential.

After reaching the bottom of the potential well (i.e. GMF1 at the end of the run, or GMF2 at the start), the clouds start to climb back up. The acceleration that the clouds feel is now acting as a ``break'' on their circular velocities, and this causes material to crowd on the downstream side of the spiral potential (see Fig.\,\ref{fig:evolution_cloudG}, for $t\geq14$\,Myrs). Clouds start to incorporate (or are being incorporated by) gas from other clouds also exiting the spiral potential, effectively creating an almost continuum of molecular gas, i.e. the gaseous spiral arm. This merging of clouds, combined with the fact that clouds often leave the potential well at an angle (i.e. feeling different accelerations along their length), results in the clouds becoming more distorted, no longer holding the imprints of their previous state as high contrast giant molecular filaments. 

In terms of chemical evolution, as the GMFs approach the spiral arm, the total mass remains relatively constant over time (Fig.\,\ref{fig:total_mass_evolution}), while the increase in density on the ridges of the filaments naturally increases the amount of both molecular gas and CO (Fig.\,\ref{fig:molecular_fraction_evolution}). Once in the gaseous spiral arm, the molecular mass fraction of the clouds still increases (Fig.\,\ref{fig:molecular_fraction_evolution}), but the molecular fraction does not reach more than $\sim$60\%. However, this limit is likely due to our prescription of the feedback, which prevents gas from reaching very high densities, which is where we would expect it to become mostly molecular.

Throughout their evolution, both GMF1 and GMF2 fragment into smaller clouds (e.g. see dashed lines on Figs.\,\ref{fig:total_mass_evolution} and \ref{fig:molecular_fraction_evolution}), which also alters their overall appearance. The reasons for clouds to break are a combination of stellar feedback and the interplay of the different forces imposed on the clouds, which we study in full detail in Sect.\,\ref{sec:forces}.

%=========================

\section{Interplay of forces}
\label{sec:forces}

\subsection{Forces on the cloud surface}
\label{sec:accel}

\begin{figure}
\centering
\includegraphics[width=0.45\textwidth]{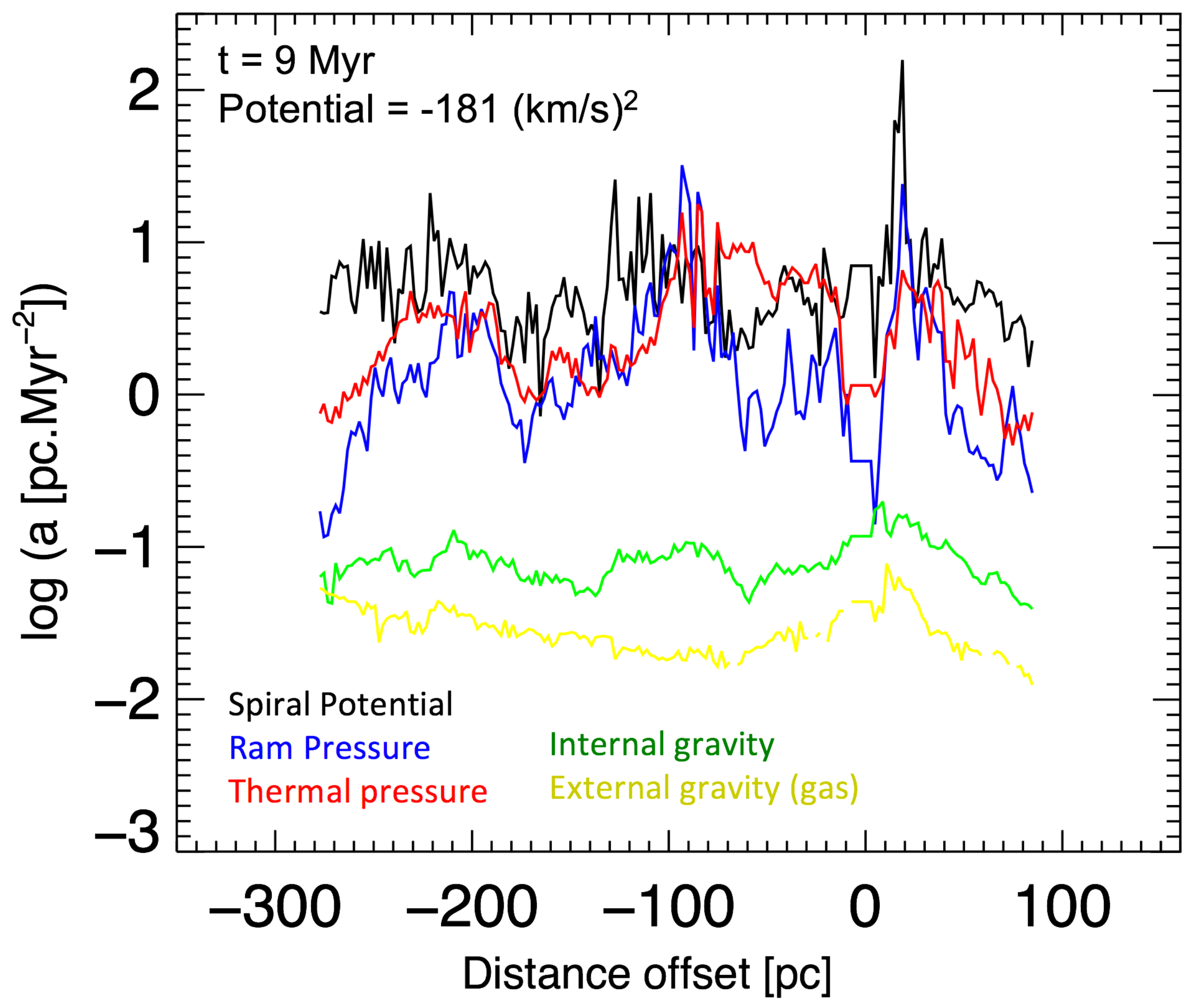}
\caption{\small{Example of the average amplitude of the accelerations estimated for the main cloud of GMF1, at $t=9$\,Myrs, as imposed by different forces on each surface point, as a function of the distance offset along the major axis direction (in pc) from a reference point on the cloud. The different coloured lines represent the acceleration from the spiral potential force (black), the ram pressure gradient (blue), the thermal pressure (red), internal gravity of the cloud (green) and external gravity from the gas around the cloud (yellow). The amplitude of the spiral potential at this particular timestep is shown on the bottom-left corner. The time evolution of these accelerations for GMF1 and GMF2, can be found in Figs.\,\ref{fig:evolution_accel_cloudA} and \ref{fig:evolution_accel_cloudG}.}}
\label{fig:accel_cloudA_t1}
\end{figure}

In order to understand what dictates the evolution of these GMFs in the simulation, we have studied the interplay of all forces on the surface of our two GMFs at 1\,pc resolution. For simplicity, given that each GMF splits up into a number of smaller clouds, we have only selected one main cloud for GMF1 and two for GMF2 to track over time (see cyan and magenta contours in Figs.~\ref{fig:evolution_cloudA} and \ref{fig:evolution_cloudG}). We used the 3D masks that defined each cloud as per our extraction algorithm to define the surface of the cloud. For each surface point, we estimate the forces that that cell feels, namely the ram pressure gradient, the thermal pressure gradient, the internal gravitational force from the gas within the cloud, the external gravitational force from the gas outside the cloud, and the gravitational force from the galactic spiral potential (see App.\,\ref{app:force_calcs} for details). We then analyse these forces in terms of their acceleration amplitude, at each surface point. An example of this is shown in Fig.\,\ref{fig:accel_cloudA_t1} (with the complete time evolution show in Figs.\,\ref{fig:evolution_accel_cloudA} and \ref{fig:evolution_accel_cloudG}), where the abscissa shows the distance offset along the major axis of the cloud (in pc) from a reference point on the cloud, and the different coloured lines shows the average amplitude of the acceleration at the respective surface points.

From the virial analysis presented in \citet{dc2016}, and also from our results shown in App.\,\ref{app:time_evolution}, we find that more than half of the clouds in these simulations, at these scales, are not gravitationally bound, with median $\alpha_{vir}$ values of $\sim$2.3 (see Table\,\ref{tab:stat_properties}). The distributions of the $\alpha_{vir}$ values (see Fig.\,\ref{fig:histograms}) have a strong tail to high values, spanning up to values in excess of 20. Such high $\alpha_{vir}$ values are not commonly observed, but this may simply be due to the fact that observations use CO to trace GMCs, and that CO, as shown in \citet[][]{dc2016}, often only traces the denser parts of larger molecular cloud complexes, hence preferably focusing on the gravitationally bound part of clouds. Although these high $\alpha_{vir}$ values would be commonly taken as suggestive that clouds may be in the process of being dispersed if not under the confinement of an external pressure, it is worth noting that such clouds, with typical velocity dispersions of the order of $\sigma_v \sim 2.5$\,km\,s$^{-1}$ (see Table\,\ref{tab:stat_properties}), would only expand by $\sim$\,25\,pc in 10 Myrs. This means that at large scales, clouds would not necessarily need external pressure to hold them together, they simply do not have enough time to expand and disperse, before something else happens to them. Even so, here we quantify the actual role of the external pressure on the evolution of our two GMFs. From Fig.\,\ref{fig:accel_cloudA_t1} (and also from Figs.\,\ref{fig:evolution_accel_cloudA} and \ref{fig:evolution_accel_cloudG}), we can see that the amplitude of the accelerations from the ram and thermal pressure forces on the surfaces of the clouds (red and blue lines respectively) are typically one to two orders of magnitude higher than that of the internal gravity (in green), which does suggest that these GMFs are mostly pressure confined. Internal gravity does play a role, within the cloud, to allow for star formation to take place, but it is not what holds these hundreds-of-parsec long structures together. The external gravity from the gas (in yellow), is typically less than that of the internal gravity, except for when nearby clouds approach the surface of the cloud studied. 

Perhaps most interestingly, when we focus on the accelerations induced by the spiral potential (in black), we can see that the amplitude of the spiral potential acceleration is orders of magnitude above that of the internal gravity, and is similar across the entire cloud (i.e. the black curve is relatively flat) for most timesteps. However, since the direction of the spiral potential force is similar along the cloud, it simply acts as a uniform force pulling the cloud as a whole towards the bottom of the potential, without significantly affecting its properties, despite its large amplitude. 
In fact, the spiral potential force is only capable of breaking and/or distorting the clouds, whenever there is a large gradient of accelerations across the cloud. This can happen, for instance, when a portion of a given cloud is closer to the bottom of the spiral potential well than the rest, creating a difference in the accelerations felt at different points on the cloud.

\subsection{What breaks the filaments?}

As star formation starts to take place within clouds, and in particular within filamentary clouds, stellar feedback is sometimes able to pierce the clouds and effectively break the giant filaments into smaller sections (see Figs.~\ref{fig:evolution_cloudA} and \ref{fig:evolution_cloudG}). For instance, GMF2 experiences a number of star formation (and thus feedback) events at several timesteps, and in several points along its length, such that they break the cloud into smaller filaments. Feedback events therefore seem to be the dominant mechanism by which GMF2 fragments. 

\begin{figure}
\centering
\includegraphics[width=0.45\textwidth]{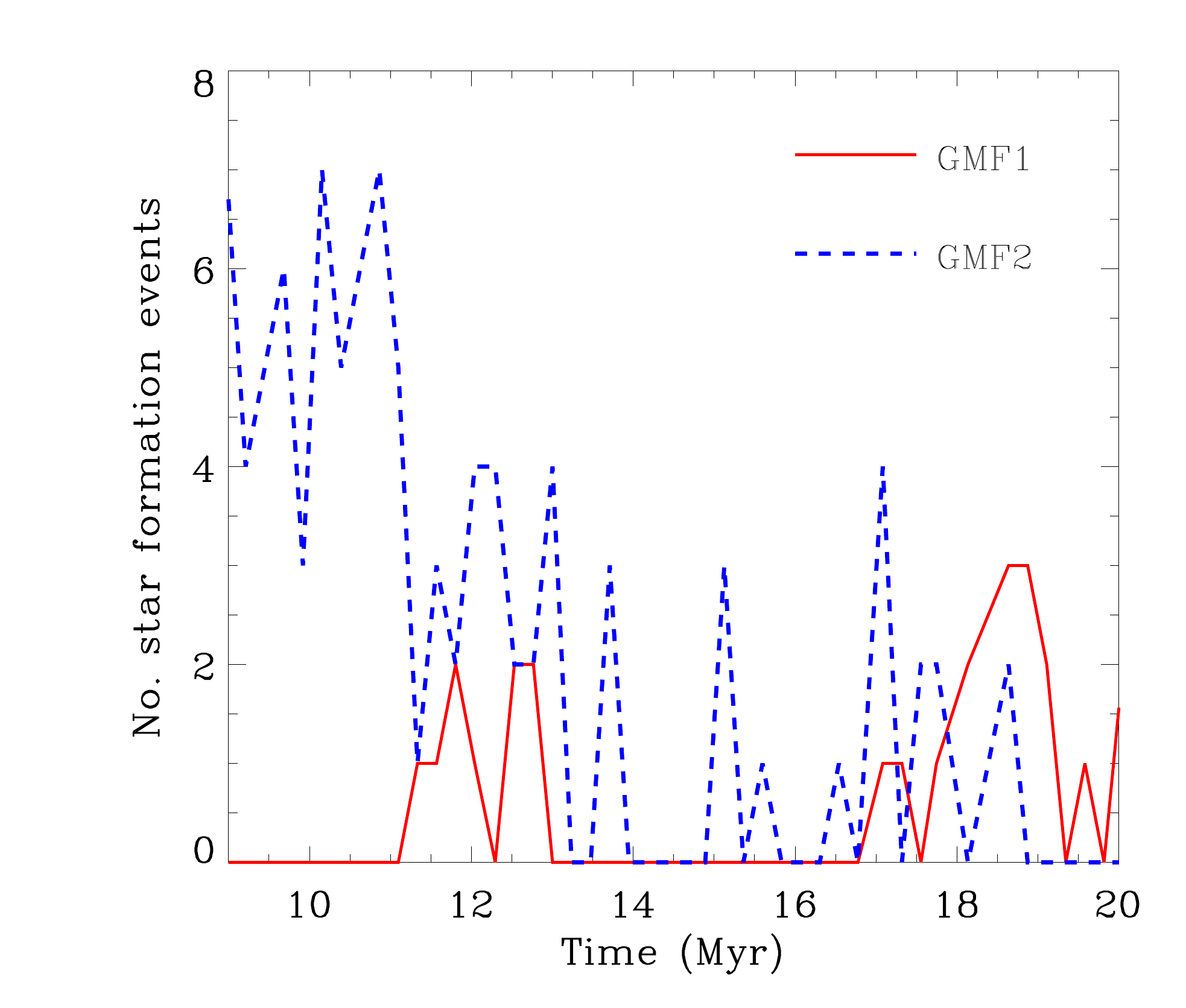}
\caption{\small{Count of number of star formation events over time, for GMF1 (solid red line) and GMF2 (dashed blue line). This was done by tracking all the gas particles that make up each GMF over time, directly on the SPH data. We can see that GMF2 is very actively forming stars between $t=9-11$\,Myrs (when it is at the bottom of the potential well), and then becomes more constant over time. GMF1 is much less active overall, but does seem to have an increase number of SF events towards the end of the simulation, when it also reaches the bottom of the potential well, at $t=17-19$\,Myrs. The late SF events of GMF1, however, are not associated with the main GMF that we track in Sect.\,\ref{sec:accel}, but instead take place within some of the smaller fragments that broke off the original filament (see bottom right panels of Fig.\,\ref{fig:evolution_cloudA}).}}
\label{fig:SF}
\end{figure}

However, this is not always the case: for instance, GMF1 suffers from a few SF and feedback events at $t=11-13$\,Mrys (see Figs.\,\ref{fig:evolution_cloudA} and \ref{fig:SF}) that are not capable of breaking the cloud. Indeed, despite its potential for disruption, stellar feedback is not the sole mechanism by which clouds are broken into smaller segments. If we inspect what happens in GMF1 (Figs.~\ref{fig:evolution_cloudA} and \ref{fig:evolution_accel_cloudA}), we can see that as the cloud approaches the bottom of the potential well at an angle, there is an increase of the spiral potential accelerations towards the centre of the cloud, with a sharp decrease towards the sides. With a variation of nearly 3 orders of magnitude in the acceleration felt at different points in the cloud, the cloud then breaks at $t=13-14$\,Myrs. Conversely, the effect of the spiral potential does not seem to be so strong for GMF2. As it starts at the bottom of the potential well, it has an almost constant gravitational pull from the potential across its length, which does not affect its shape. The filament then starts to climb up the potential well at an angle with respect to the spiral arms, and does sometimes feel large variations in the accelerations due to the potential. However, often these are of smaller amplitude than those due to the external ram and thermal pressure (and sometimes even below those from the internal and external gravity forces), and so the spiral potential is insufficient to break up the cloud.

%============================================

\section{Discussion}
\label{sec:discussion}

\subsection{No `bones' of the galaxy?}
Previous observational studies have suggested that the GMFs in the Galaxy form some sort of skeleton, or ``bones of the Galaxy'', tracing the spiral arms \citep{Goodman2014,Zucker2015,Wang2016}. This contrasts somewhat with observations of external galaxies, where there are clear filamentary structures in the inter-arm regions, but in the spiral arms, the gas tends to exhibit a more complex morphology. Our results are clearly more consistent with the observations of external galaxies. Giant molecular filaments exist everywhere in the inter-arm regions, but they are simply sheared clouds of gas, and do not constitute a physically relevant structure. Furthermore they cease being giant filaments once they become or join the giant molecular clouds that make up the gaseous spiral arms. 

Although striking in their shape and form, giant molecular filaments are not gravitationally bound clouds as a whole. These are structures that are naturally shaped and confined by their global galactic environment (shaped by the rotational shear, and confined by the external pressure), and thus they are unsurprisingly coherent in velocity. We caution, however, that this should not be interpreted as proof that these are physically relevant structures. For instance, we find that GMFs often break before they reach the gaseous spiral arm, and the resulting smaller filamentary sections maintain their relative position and motion, particularly up until they cross the bottom of the potential well. This means that these multiple filaments retain their apparent alignment, despite effectively being disconnected from each other in density space, no longer representing a long and single ``coherent structure" per se. This naturally explains why observations of GMFs in our Galaxy typically also find a number of ``break" points along GMFs, with other filaments starting further out, but still retaining the same velocities and apparent alignment \citep[e.g. the ``Nessie extended'' filament,][]{Goodman2014}. The physical relevance of such structures, however, should be interpreted with caution.

The results from our models suggest that the extremely long filaments might preferably trace the regions of entry into the spiral arms. This idea is still consistent with the observational data for our Galaxy, especially considering the large uncertainties associated with the determination of kinematical distances for Galactic observations, which are greatest for clouds near spiral arms where the velocities deviate from their circular motions. There, the distance uncertainties are of the order of the arm width itself, making it incredibly hard to discriminate between a cloud in a spiral arm and one close to it, solely based on kinematical information.

%=================

\section{Summary and Conclusions}

In this paper, we have studied the time evolution of molecular clouds (and giant molecular filaments in particular) in a high-resolution section of a spiral galaxy simulation, over a period of 11\,Myrs. Our main results can be summarised as follows:
\begin{itemize}
\item The statistical properties of clouds at a given snapshot are representative of the overall trends of evolution of cloud properties over time.
\item Giant molecular filaments are only found in inter-arm regions, in accordance with the results suggested by \cite{dc2016}. They are formed from gas clouds that exit from an arm (typically nearly perpendicularly to it), enter the shear-dominated inter-arm region, and get stretched by the differential rotation of the gas. 
GMFs become more well defined and well aligned with the spiral arms' axis as they approach the bottom of the spiral potential. 
\item An analysis of the balance of forces acting on the surface of these GMFs shows that these clouds are not gravitationally bound as a whole, but are pressure confined (by ram and thermal pressure) to at least some extent. The gravitationally bound part of these clouds is only confined to smaller, local higher density regions within the clouds where SF takes place, best traced by CO.
\item The gas within GMFs is forming stars before entering the gaseous spiral arm, and, at least in our selected sample, there is no particular increase of star formation events once they enter the spiral arm. 
\item These giant filaments get broken into smaller filament sections over time (some still as long as $\sim$100\,pc), either due to the star formation and the subsequent stellar feedback breaking the cloud, or due to the differential force from the gravitational potential on different portions of the cloud.
\item When GMFs enter the gaseous spiral arm, they incorporate/are incorporated into the more continuous medium that makes up the gaseous spiral arm. This changes the morphology of clouds, from high-contrast elongated structures, to become more sub-structured GMC complexes. 
\end{itemize}

Both from the statistical analysis of the position of GMFs in the galaxy \citep[from][]{dc2016}, and the analysis of the time evolution of GMFs presented here, our results strongly suggest that GMFs are not able to survive crossing the gaseous spiral arm, which implies that they do not trace the spiral structure of galaxies, as suggested by some observational studies of our Galaxy. For Galactic studies, however, we caution that the kinematical distances around spiral arms are highly uncertain, because the spiral arms deviate from the circular motions of the gas. The high uncertainties related with de-projecting these filaments from a PPV into a PPP perspective, should therefore be taken into account when trying to associate clouds with spiral arms. We find that GMFs are most pronounced when they reach the bottom of the potential well (i.e. just before entering the spiral arm), which could account for the fact that, observationally, we cannot distinguish if these clouds are inside the spiral arms, or just before/after.

There are, however, a few caveats associated with our study, which could affect our results. For instance, with our specific implementation of stellar feedback, we do not allow clouds to reach very high densities, and therefore do not resolve the details of SF occurring within the clouds. This could potentially have an impact on how clouds evolve and how stellar feedback may or may not be able to break clouds apart. These models also do not yet include magnetics fields, which could perhaps change how these structures evolve over time. For a high number of spiral arms, or a different rotation curve, the amount of time that a cloud spends in the inter-arm regions, and the shear it experiences, may differ, altering how the giant filaments get stretched, their resultant shapes, and how well they become aligned with the approaching spiral arms. Finally, in these models, the underlying spiral potential is of a ``grand design'' form, in that there is a fixed pattern speed, and the gas flows through the arms (up and down the spiral potential). Using a more realistic live stellar potential (as opposed to a fixed potential) could also produce different results as the dynamics of the gas across the spiral arms may be different \citep[e.g.][]{Pettitt2015,Baba2017}. Indeed, such arms are more dynamic and tend to be transient, in which case the response of the gas might not be the same as what we find here, and the formation of spurs may not occur in the same way by shearing of spiral arm GMCs \citep[though see][for spur formation in interacting galaxies]{Pettitt2016}. Although outside the scope of this paper, it would be interesting to explore if the ability to form giant filaments of the kind that are supposedly observed (and their survival) as a function of the specific type of spiral arms.

\section*{Acknowledgements}

We thank the anonymous referee whose comments and suggestions have helped making the paper clearer. We would also like to thank J. E. Pringle and S. Clarke for numerous insightful discussions. ADC and CLD acknowledge funding from the European Research Council for the FP7 ERC starting grant project LOCALSTAR. ADC also acknowledges the support of the UK STFC consolidated grant ST/N000706/1. This work used the DiRAC Complexity system, operated by the University of Leicester IT Services, which forms part of the STFC DiRAC HPC Facility (www.dirac.ac.uk ). This equipment is funded by BIS National E-Infrastructure capital grant ST/K000373/1 and  STFC DiRAC Operations grant ST/K0003259/1. DiRAC is part of the National E-Infrastructure. This work also used the University of Exeter Supercomputer, a DiRAC Facility jointly funded by STFC, the Large Facilities Capital Fund of BIS, and the University of Exeter.

%%%%%%%%%%%%%%%%%%%%%%%%%%%%%%%%%%%%%%%%%%%%%%%%%%



%\bibliographystyle{mnras}

\bibliographystyle{mnras}
\bibliography{fb}

%%%%%%%%%%%%%%%%%%%%%%%%%%%%%%%%%%%%%%%%%%%%%%%%%%

%%%%%%%%%%%%%%%%% APPENDICES %%%%%%%%%%%%%%%%%%%%%

\appendix
\section{Global time evolution}
\label{app:time_evolution}

\begin{figure*}
\centering
\hspace{-0.2cm}
\includegraphics[width=0.33\textwidth]{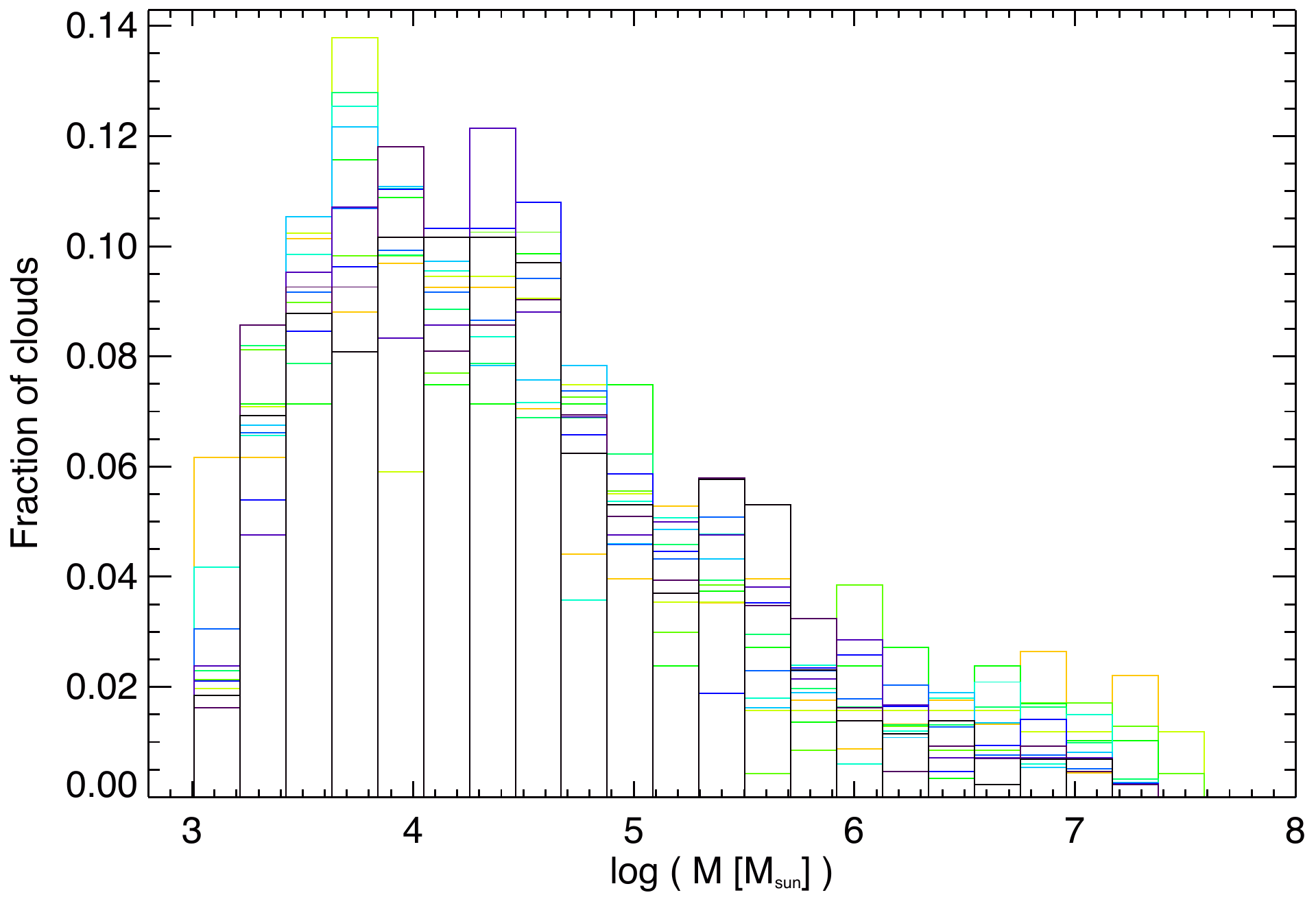}
\includegraphics[width=0.33\textwidth]{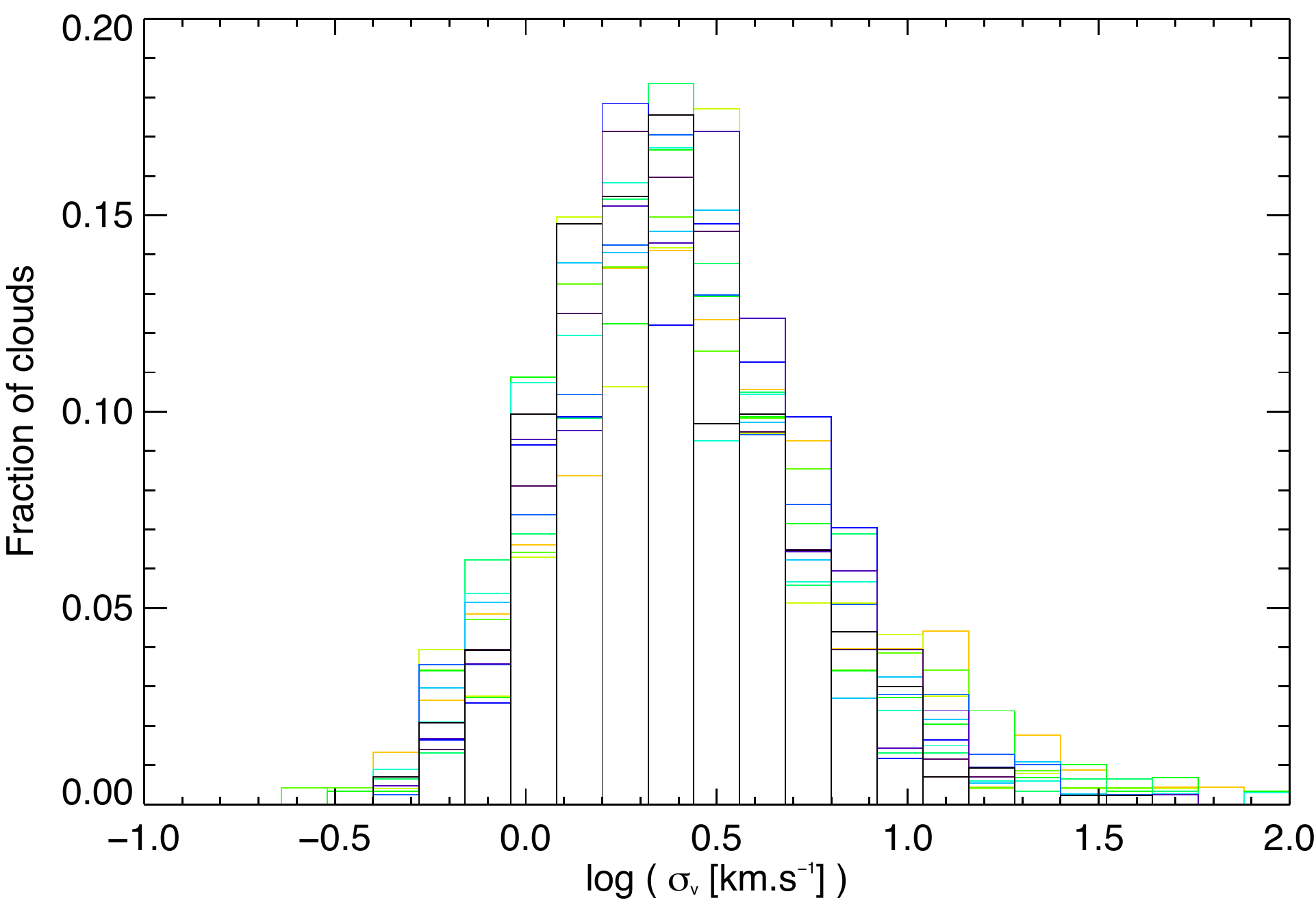}
\includegraphics[width=0.33\textwidth]{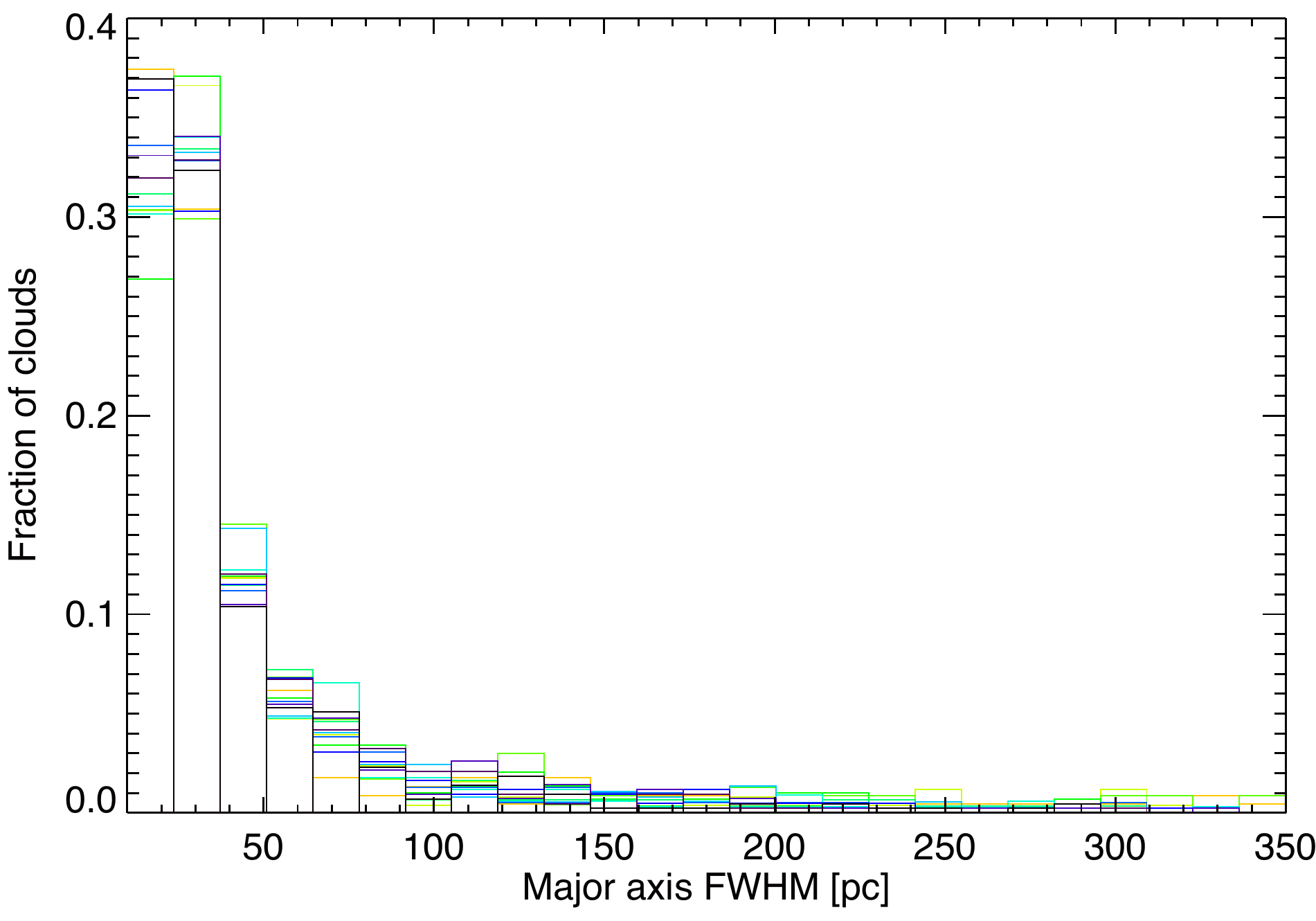}\\
\hspace{-0.2cm}
\includegraphics[width=0.33\textwidth]{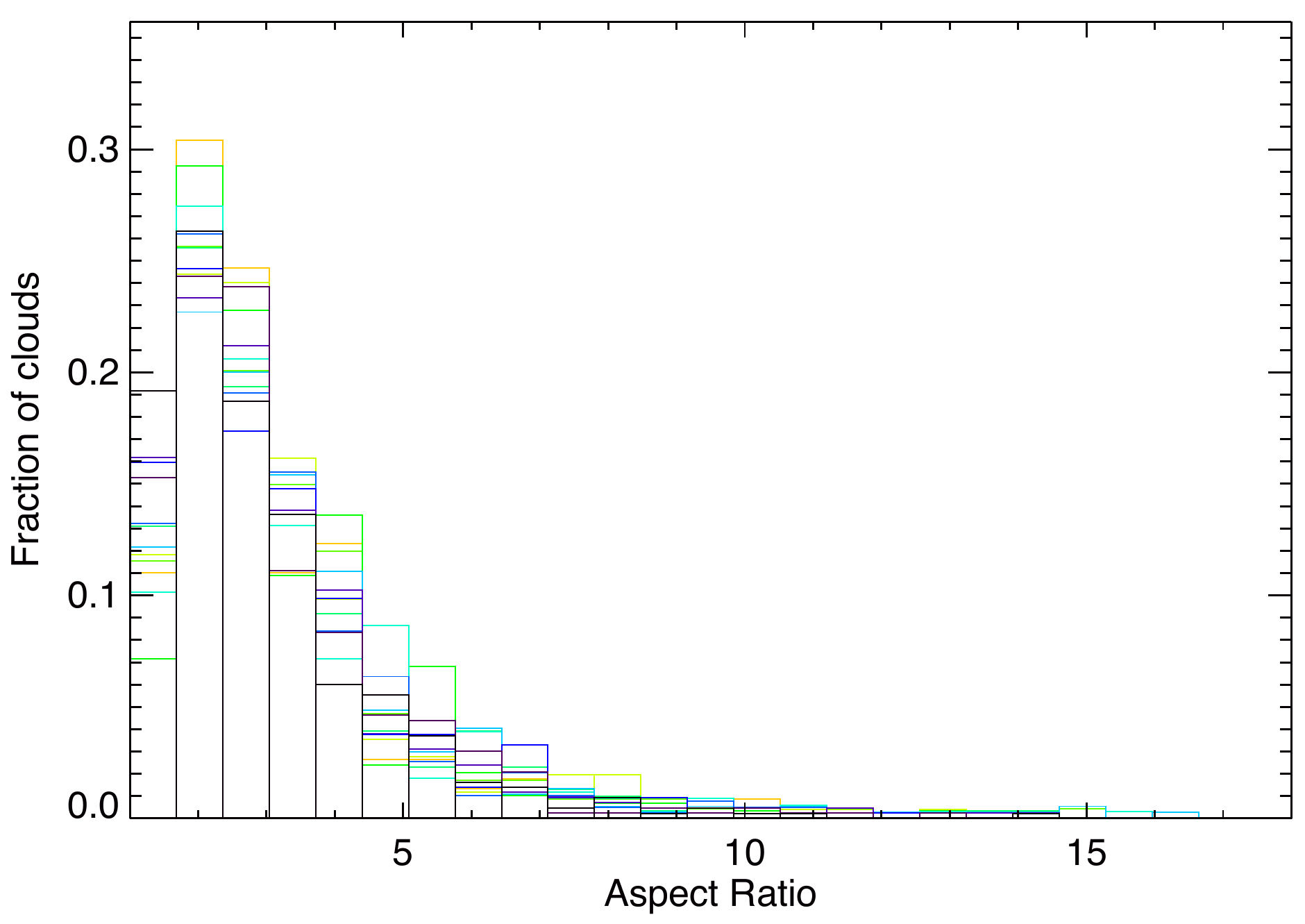}
\includegraphics[width=0.33\textwidth]{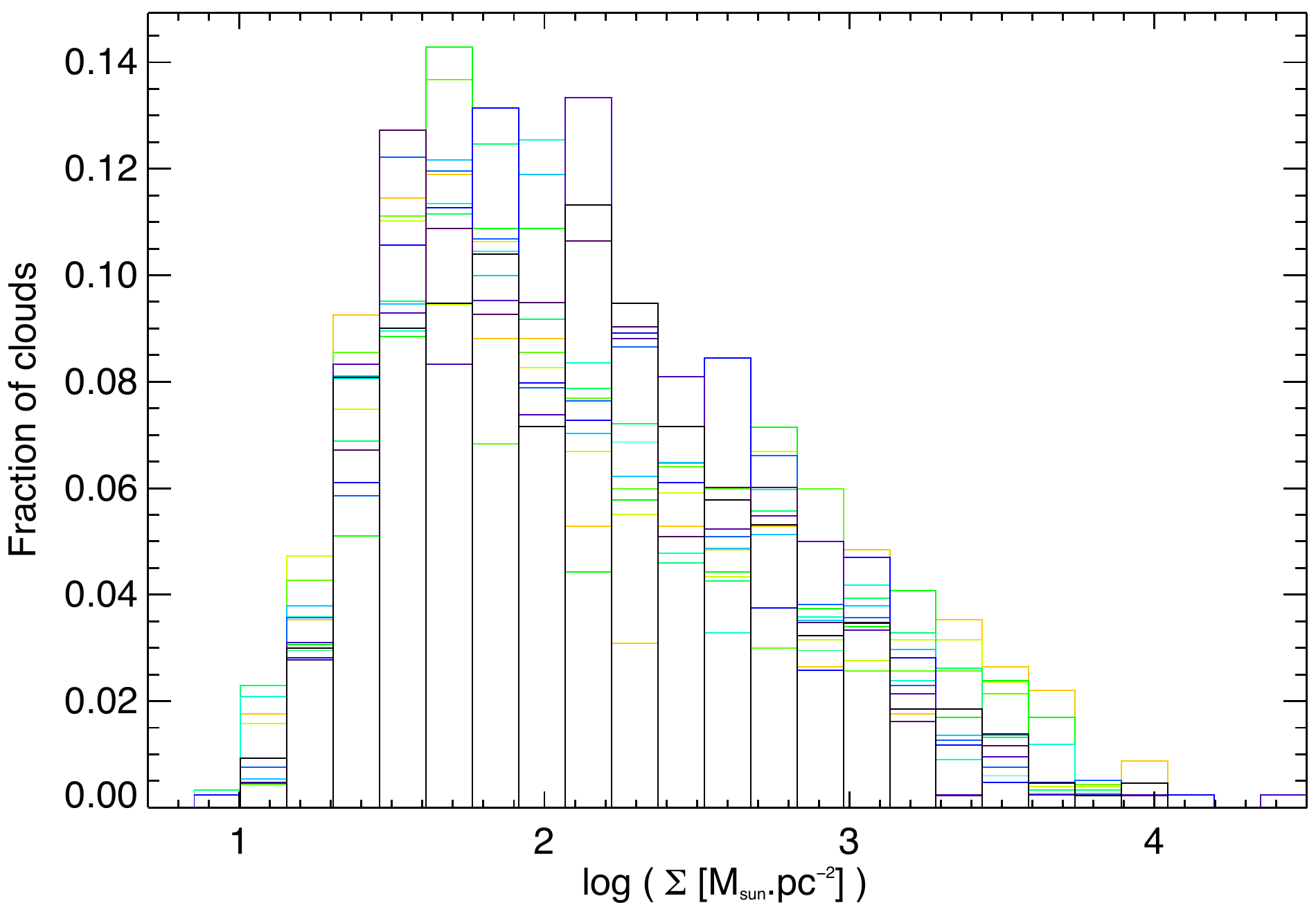}
\includegraphics[width=0.33\textwidth]{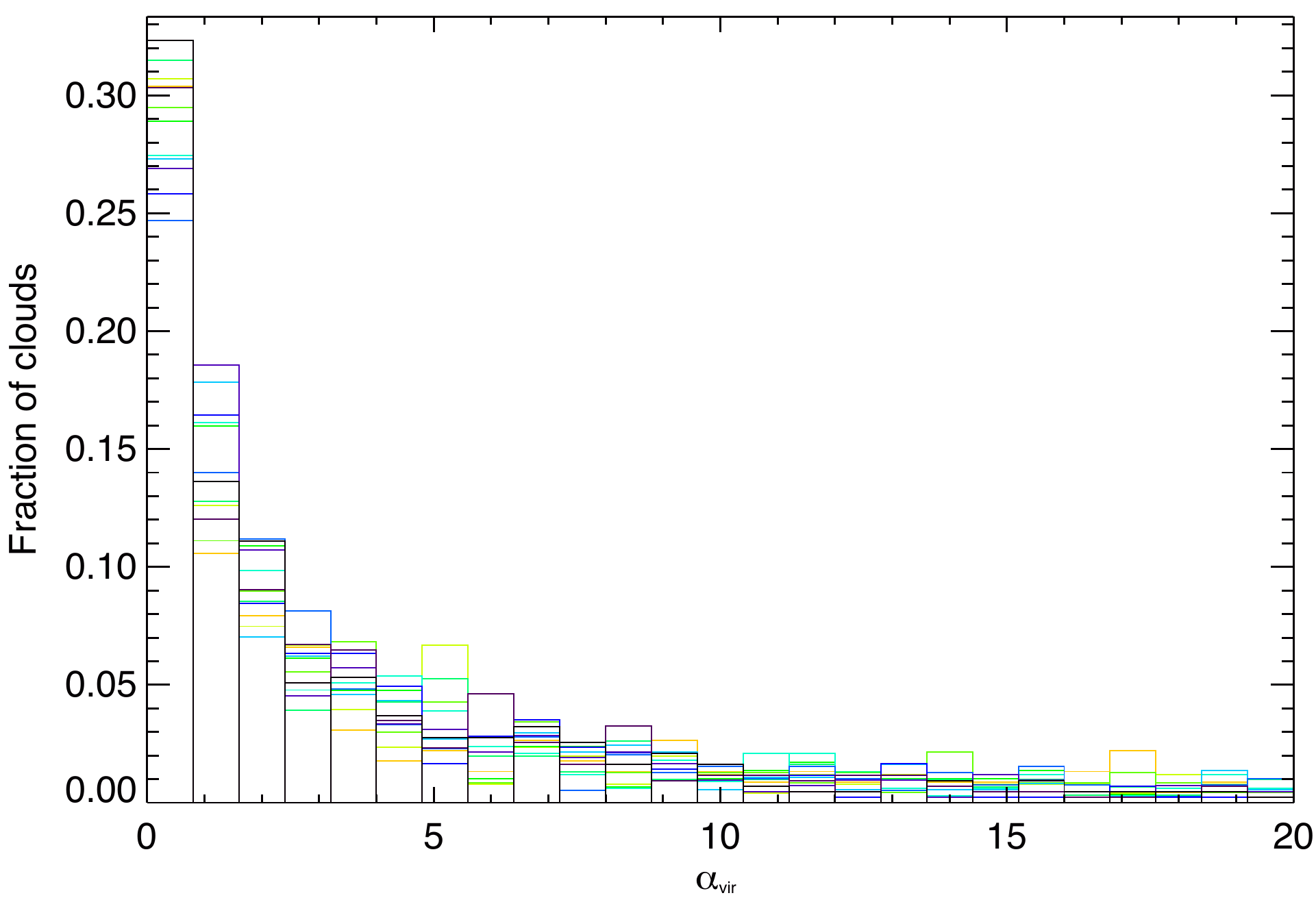}
\caption{\small{Histograms of the global cloud properties, whose statistics are shown in Table\,\ref{tab:stat_properties}: Mass (top-left), velocity dispersion (top-centre), major axis (top-right), aspect ratio (bottom-left), surface density (bottom-centre), and virial parameter (bottom-right). The histograms for the 12 timesteps are shown as different colours, each of which is normalised by the total number of clouds in that particular timestep.}}
\label{fig:histograms}
\end{figure*}

We have studied the global time evolution of clouds in our simulation using the {\sc{scimes}} code \citep{Colombo15} on 3-dimensional datacubes of the H$_{2}$ density,  with 5\,pc-size cells. The position and extent of the GMCs extracted at each timestep can be seen in Figs.\,\ref{fig:top-down-molecular} to \ref{fig:top-down-molecular2}, as seen with a face-on (top-down) perspective of the simulated galaxy. We estimated a number of properties for the clouds, namely their H$_{2}$ mass ($M_{\rm H_{2}}$), velocity dispersion ($\sigma_{v}$), major axis FWHM (in 3D), aspect ratio ($A_{r}$, computed as the ratio of the major axis to the average of the other two axes), average surface density ($\Sigma$, calculated taking the area of a circle with the equivalent radius of the cloud) and virial parameter ($\alpha_{vir} = 5 \sigma_v^{2}R / GM$). The shape of the distributions are similar on all timesteps (see Fig.\,\ref{fig:histograms}), and the median values, estimated for each timestep and as a temporal average, are shown in Table\,\ref{tab:stat_properties}.

%\onecolumn
\begin{table*}
\caption{Statistical properties of GMCs from the H$_{2}$ densities over time. The values presented correspond to the median value of each property at each timestep, and the dispersions were derived as the mean value of the absolute deviation of the first and third quartiles from the median.}
\label{tab:stat_properties}
\renewcommand{\footnoterule}{}  
\begin{tabular}{l c c c c c c}
\hline 
\hline
Time		&  $log (M_{\rm H_{2}})$		&	$log (\sigma_{v})$	&	Major axis		& 	$A_{r}$	&	 $log (\Sigma) $ &	 $\alpha_{vir}$	\\
(Myrs)	& log (M$_{\odot}$)	&	log (km\,s$^{-1}$)	&	(pc)			&			&	(M$_{\odot}$\,pc$^{-2})$	&				\\
\hline 
\hline
9		&	$4.24 \pm  0.67$	& $0.44 \pm 0.24$	& $26.1 \pm 10.8$	& $2.5 \pm 0.8$ & $1.97 \pm  0.56$ 	& $2.5 \pm   5.6$	\\
10 		&      $4.29 \pm  0.60$	& $0.41 \pm 0.22$	& $28.7 \pm 11.1$	& $2.7 \pm 0.8$ & $2.01 \pm  0.51$ 	& $2.3 \pm   3.6$ 	\\
11 		&      $4.35 \pm  0.59$	& $0.38 \pm 0.22$	& $30.5 \pm 14.5$	& $2.7 \pm 0.9$ & $2.00 \pm  0.47$ 	& $2.4 \pm   3.5$ 	\\
12 		&      $4.36 \pm  0.60$	& $0.39 \pm 0.23$	& $29.7 \pm 13.4$	& $2.8 \pm 0.9$ & $2.05 \pm  0.49$ 	& $2.0 \pm   2.8$ 	\\
13 		&      $4.26 \pm  0.63$	& $0.38 \pm 0.20$	& $30.1 \pm 13.7$	& $2.8 \pm 0.9$ & $1.98 \pm  0.46$ 	& $2.2 \pm   3.5$ 	\\
14 		&      $4.15 \pm  0.63$	& $0.35 \pm 0.22$	& $29.7 \pm 13.4$	& $2.7 \pm 0.9$ & $1.97 \pm  0.41$ 	& $2.0 \pm   2.9$ 	\\
15 		&      $4.22 \pm  0.58$	& $0.36 \pm 0.22$	& $29.5 \pm 12.6$	& $2.9 \pm 0.9$ & $1.99 \pm  0.41$ 	& $2.1 \pm   3.1$ 	\\
16 		&      $4.28 \pm  0.60$	& $0.38 \pm 0.21$	& $29.3 \pm 13.0$	& $2.7 \pm 0.9$ & $2.01 \pm  0.43$ 	& $2.4 \pm   3.7$ 	\\
17 		&      $4.32 \pm  0.53$	& $0.41 \pm 0.21$	& $28.1 \pm 11.9$	& $2.6 \pm 0.9$ & $2.02 \pm  0.43$ 	& $2.4 \pm   3.0$ 	\\
18 		&      $4.37 \pm  0.60$	& $0.40 \pm 0.20$	& $28.4 \pm 13.5$	& $2.7 \pm 0.9$ & $2.11 \pm  0.40$ 	& $1.9 \pm   2.9$ 	\\
19 		&      $4.29 \pm  0.60$	& $0.37 \pm 0.18$	& $29.6 \pm 12.4$	& $2.6 \pm 0.9$ & $2.03 \pm  0.42$ 	& $2.4 \pm   2.9$ 	\\
20 		&      $4.33 \pm  0.58$	& $0.34 \pm 0.21$	& $28.0 \pm 11.5$	& $2.5 \pm 0.8$ & $2.10 \pm  0.40$ 	& $1.9 \pm   2.7$ 	\\
\hline 
Temporal Median$^{(a)}$ & $4.29 \pm 0.04$ & $0.38 \pm 0.02$	& $29.5 \pm 0.6$	& $2.7 \pm 0.1$ & $2.01 \pm 0.03$ 	& $2.3 \pm  0.2$	\\
\hline
\end{tabular}
\flushleft
$^{(a)}$ The quoted dispersions refer to the mean deviation of the absolute median values of each property at different times, with respect to the global temporal median (i.e. it does not incorporate the variations within each timestep).\\
\end{table*}

\section{Force calculations}
\label{app:force_calcs}

To understand the role of different forces in shaping the GMFs over time, we have estimated, for each point at the surface of the clouds, the contributions of the forces from the internal gravity, external gravity from the gas, external gravity as imposed by the spiral potential, thermal pressure gradient, and ram pressure gradient. We then compare the amplitude of the accelerations imposed by these forces on each parcel of gas at the surface of the cloud. In this section we detail how the different forces are estimated for each surface point. For these calculations, we define $\vec{r}$ as the 3D positional vector of a cell, and $\vec{v}$ its 3D velocity vector, such that if we consider two cells $\rm{i}$ and $\rm{j}$, then $\vec{r}_{\rm{ij}}=\vec{r}_{\rm{j}} - \vec{r}_{\rm{i}}$, and $\vec{v}_{\rm{ij}} = \vec{v}_{\rm{j}}-\vec{v}_{\rm{i}}$.

\subsection{Gravitational force from the gas}

The 3-dimensional gravitational force vector between any two cells, $\rm{i}$ and $\rm{j}$, can be described by
\begin{equation}
\vec{F}_{\rm{g}}= \frac{G m_{\rm{i}} m_{\rm{j}}}{d^{2}}  \hat{r}_{\rm{ij}},
\end{equation}
where $G$ is the gravitational constant, $m_{\rm{i}}$ and $m_{\rm{j}}$ are the total masses of cells $\rm{i}$ and $\rm{j}$ respectively,  $d$ is the distance between the two cells, and $\hat{r}_{\rm{ij}}$ is the positional unity vector between $\rm{i}$ and $\rm{j}$. We then separate the contribution from the internal gravity of the cloud and that of the surrounding gas. For the internal gravitational force, $\vec{F}_{\rm{g}}^{\rm{int}}$, we have estimated the total gravitational force vector of each surface point using only cells within the cloud. For the external gravitational force, $\vec{F}_{\rm{g}}^{\rm{ex}t}$, we estimated the total gravitational force vector of each surface point using all the gas external to the cloud, situated up to a distance of 100\,pc from the cloud.

\subsection{Ram pressure force}

To estimate the ram pressure at each surface point $\rm{i}$ we used only the immediate neighbouring cells $\rm{j}$, i.e. within 1\,pc radii. For the ram pressure, only the component of the velocity along $\vec{r}_{\rm{ij}}$ is relevant, and thus we define $\vec{v}_{r}$ as:
\begin{equation}
\vec{v_{r}} = \frac{(\vec{v}_{\rm{ij}} . \vec{r}_{\rm{ij}})}{|\vec{r}_{\rm{ij}}|} \hat{r}_{\rm{ij}}.
\end{equation}
The total ram pressure vector, felt at the surface point ${\rm{i}}$, can then be expressed as the vectorial sum of the ram pressure from all points $\rm{j}$ that surround cell $\rm{i}$, as 
\begin{equation}
\vec{P}_{\rm{ram}}= \sum_{j}  \rho_{\rm{j}} |\vec{v}_{r}|^{2} \hat{r}_{\rm{ij}}.
\end{equation}
The ram pressure force, can then be expressed as
\begin{equation}
\vec{F}_{\rm{ram}}=\vec{P}_{\rm{ram}} S,
\end{equation}
where $S$ is the surface of the cell that feels the pressure (in our case, $S=1$\,pc$^{2}$).

\subsection{Thermal pressure force}

The thermal pressure of each cell is estimated as $P_{\rm{Th}} = \rho k_b T$, where $\rho$ is the total density of the cell, $T$ its temperature, and $k_b$ is the Boltzmann constant. With this, we can calculate the total pressure gradient vector, $\nabla\vec{P}_{\rm{Th}}$, between each cloud's surface cell $\rm{i}$, and all its immediate neighbours $\rm{j}$ (i.e. within 1\,pc radii, as was done for the ram pressure). In practice, for each cell $\rm{i}$, the thermal pressure gradient was estimated as
\begin{equation}
\nabla\vec{P}_{\rm{Th}} = \sum_{j} \Big(- \frac{(P_{\rm{j}} - P_{\rm{i}})}{\delta x_{\rm{ij}}}, - \frac{(P_{\rm{j}} - P_{\rm{i}})}{\delta y_{\rm{ij}}}, - \frac{(P_{\rm{j}} - P_{\rm{i}})}{\delta x_{\rm{ij}}} \Big).
\end{equation}
The thermal pressure force, can then be described as 
\begin{equation}
\vec{F}_{\rm{Th}} = \nabla\vec{P}_{\rm{Th}} V,
\end{equation}
where $V$ is the volume on which the pressure gradient was estimated (in our case, $V=1$\,pc$^{3}$). 

\subsection{Spiral potential force}

The analytic expression of the spiral potential $\psi$ \citep[from][]{DobbsBP2006}, is given by
\begin{equation}
\begin{split}
\psi_{sp}(r,\theta,t)&=-4\pi G H \rho_{0} \exp(-\frac{r-r_0}{R_s}) \sum_{n=1}^{3}
\frac{C_n}{K_n D_n} \cos(n\gamma) \\
\text{where} \quad
\gamma&=N\bigg[\theta-\Omega_p t-\frac{\ln(r/r_0)}{\tan(\alpha)}\bigg], \\
K_n&=\frac{nN}{r \sin(\alpha)}, \\
D_n&=\frac{1+K_n H +0.3(K_n H)^2}{1+0.3K_nH}, \\
C(1)&=8/(3\pi), \quad C(2)=1/2, \quad C(3)=8/(15\pi).
\end{split}
\end{equation}
The number of arms is given by N, which here is 2 for a 2-armed spiral potential. The pitch angle is $\alpha=15^o$, the amplitude of the perturbation is $\rho_0=1$~atom~cm$^{-3}$ (which leads to an amplitude of the potential of $\approx 200$~km$^2$~s$^{-2}$), and the pattern speed is $\Omega_p = 2 \times 10^{-8}$~rad~yr$^{-1}$ (which leads to a co-rotation radius of 11~kpc). We also take $r_0=8$~kpc, $R_s=7$~kpc and $H=0.18$~kpc, which are radial parameters similar to the Milky Way. 

The gravitational energy, $U$, of each cell $\rm{i}$, at a timestep $t$, due to the spiral potential, is given by
\begin{equation}
U_{\rm{i}}=\psi_{\rm{i}}(r_{\rm{i}},\theta_{\rm{i}}) m_{\rm{i}}.
\end{equation}
The gravitational potential force from the spiral arms, $\vec{F}_{\rm{pot}}$, can then be described, in cylindrical coordinates, as
\begin{equation}
 \vec{F}_{\rm{pot}} = \Big( \frac{dU_{\rm{i}}}{dr}, \frac{dU_{\rm{i}}}{d\theta}, 0 \Big) = m_{\rm{i}} \Big( \frac{d\psi_{\rm{i}}}{dr}, \frac{d\psi_{\rm{i}}}{d\theta}, 0 \Big)
\end{equation}
This can then be easily reconverted back onto cartesian coordinates, to obtain the spiral potential force vector in the same referential as the other forces.

\subsection{From forces to accelerations}

\begin{figure*}
\includegraphics[width=\textwidth]{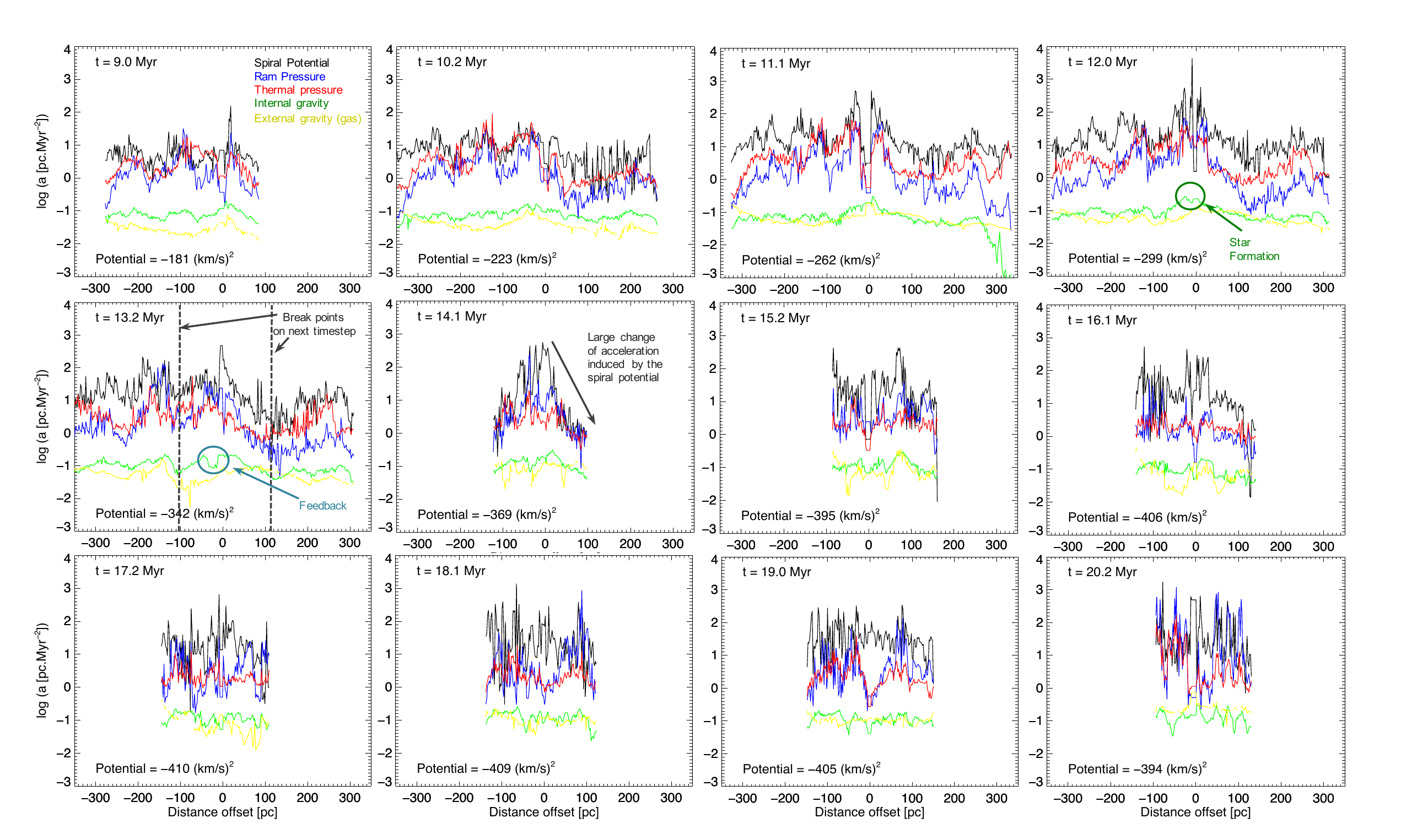}
\vspace{-0.2cm}
\caption{\small{Time evolution of the accelerations felt on the surface of the cloud, calculated for the main cloud of GMF1. Each panel shows the average amplitude of the accelerations imposed by different forces on each surface point, as a function of the distance offset along the major axis direction (in pc) from a reference point on the clouds. The different coloured lines show the acceleration from the spiral potential force (black), the ram pressure (blue), the thermal pressure (red), internal gravity of the cloud (green) and external gravity from the gas around the cloud (yellow). The amplitude of the spiral potential for each timestep is shown on the bottom-left corner of each panel. Some interesting features (e.g. star formation, stellar feedback, and breaking points of the clouds), are highlighted at the relevant timesteps. }}
\label{fig:evolution_accel_cloudA}
\end{figure*}

\begin{figure*}
\includegraphics[width=\textwidth]{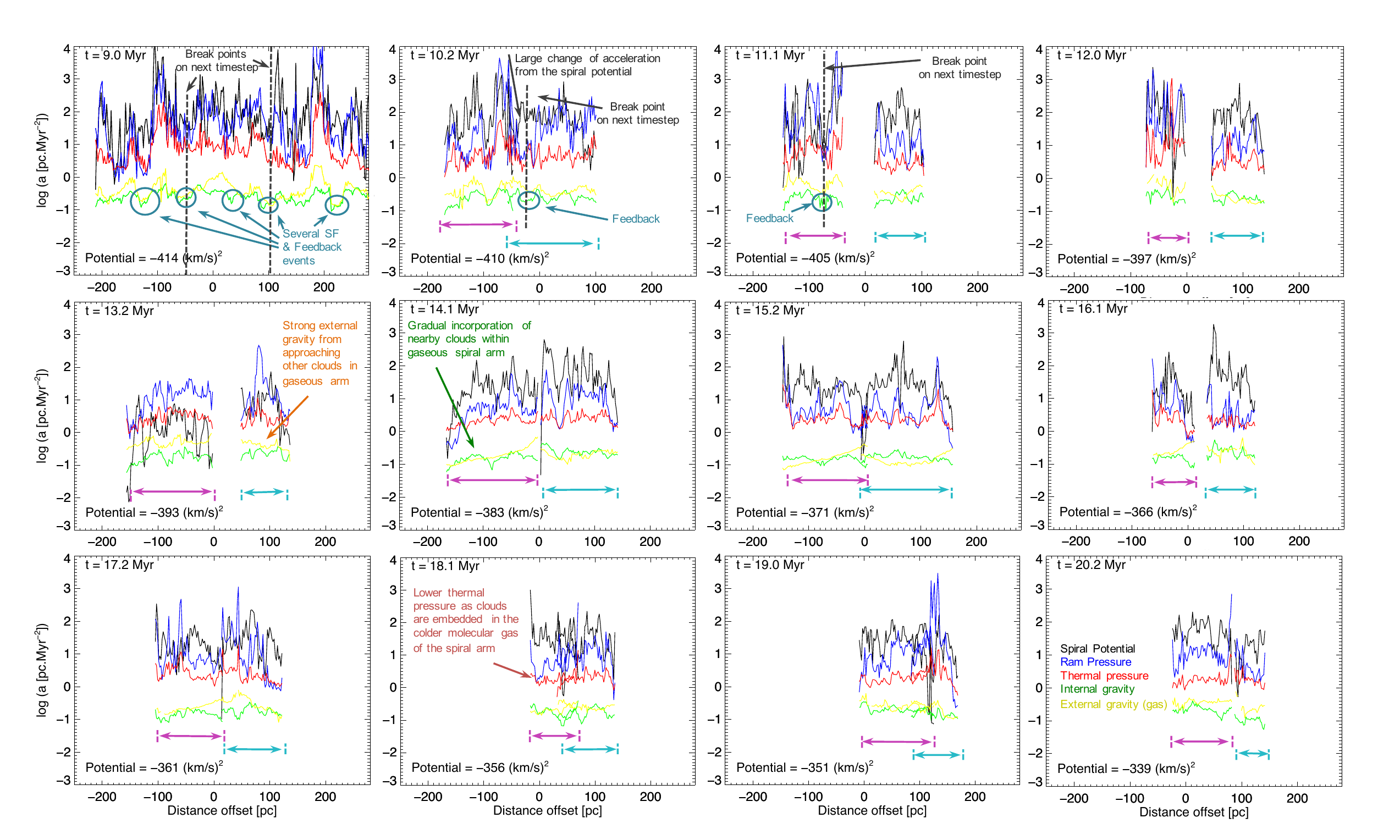}
\vspace{-0.2cm}
\caption{\small{Time evolution of the accelerations felt on the surface of GMF2. Colours, labels and lines are the same as in Fig.\,\ref{fig:evolution_accel_cloudA}. For GMF2, given that we track two clouds over time (and that these can sometimes overlap in the y coordinate), we highlight their extent as magenta and cyan bars at the bottom of each panel, so as to match the contouring of the respective clouds in Fig.\,\ref{fig:evolution_cloudG}.}}
\label{fig:evolution_accel_cloudG}
\end{figure*}

To study the interplay of the different forces across the entire cloud, it is more useful to use accelerations, so that we lose the dependency on the mass of the cell on which the forces are calculated. This is more intuitive to understand and compare. Therefore, for each surface point ${\rm i}$, we estimate the accelerations that the different forces are capable of inducing on the gas, as $\vec{a}=\vec{F}/m_{\rm{i}}$. Figures \ref{fig:evolution_accel_cloudA} and \ref{fig:evolution_accel_cloudG} show the average amplitude of these different accelerations on the surface of the giant filaments, as a function of the distance along the direction of the major axis. To help the reading of these figures, some of the main events that occur during the time evolution of the clouds are highlighted.

%%%%%%%%%%%%%%%%%%%%%%%%%%%%%%%%%%%%%%%%%%%%%%%%%%

% Don't change these lines
\bsp	% typesetting comment
\label{lastpage}
\end{document}